\documentclass[letterpaper,usegraphicx,useAMS,onecolumn,usenatbib]{mn2e}

\usepackage{amsmath,amssymb}

\def\cm{\hbox{\;cm}}
\def\yr{\hbox{\;yr}}

\def\bfc{{\bf c}}
\def\bfe{\hat{\bf e}}

\def\half{{\textstyle{\frac{1}{2}}}}

\def\bfL{{\bf L}}
\def\bfn{\hat{\bf n}}

\def\bfa{{\bf a}}

\def\bfb{{\bf b}}

\def\cross{\mbox{\boldmath $\times$}}

\def\half{{\textstyle{1\over2}}}

\def\p{\upartial}

\def\bfC{{\bf C}}

\def\bfT{{\bf T}}

\def\bfC{{\bf C}}
\def\comment#1{}

\def\gothr{\,\mathfrak R}

\newcommand{\mstar}{ M_{\star}}
\newcommand{\rstar}{ r_{\star}}
\newcommand{\rsun}{\;{\rm R}_{\odot}}
\newcommand{\msun}{\;{\rm M}_{\odot}}
\newcommand{\mbh}{M_{\bullet}}

\title{Dynamics of warped accretion discs}

\author[S.\ Tremaine and S.\ W.\ Davis]{ Scott
  Tremaine$^1$\thanks{E-mail: tremaine@ias.edu}, Shane W.\
  Davis$^2$\thanks{E-mail: swd@cita.utoronto.ca} 
\\ $^1$Institute for Advanced Study,
   Princeton, NJ 08540, USA \\
$^2$Canadian Institute for Theoretical Astrophysics,
  University of Toronto, Toronto, ON M5S 3H8, Canada}

\begin{document}

\maketitle

\begin{abstract}
  Accretion discs are present around both stellar-mass black holes in
  X-ray binaries and supermassive black holes in active galactic
  nuclei. A wide variety of circumstantial evidence implies that many
  of these discs are warped. The standard Bardeen--Petterson model
  attributes the shape of the warp to the competition between
  Lense--Thirring torque from the central black hole and viscous
  angular-momentum transport within the disc. We show that this
  description is incomplete, and that torques from the companion star
  (for X-ray binaries) or the self-gravity of the disc (for active
  galactic nuclei) can play a major role in determining the properties
  of the warped disc. Including these effects leads to a rich set of
  new phenomena. For example, (i) when a companion star is present and
  the warp arises from a misalignment between the companion's orbital
  axis and the black hole's spin axis, there is no steady-state
  solution of the Pringle--Ogilvie equations for a thin warped disc
  when the viscosity falls below a critical value; (ii) in AGN
  accretion discs, the warp can excite short-wavelength bending waves
  that propagate inward with growing amplitude until they are damped
  by the disc viscosity. We show that both phenomena can occur for
  plausible values of the black hole and disc parameters, and briefly
  discuss their observational implications.
\end{abstract}

\begin{keywords}
accretion, accretion discs -- black hole physics -- hydrodynamics --
binaries: close -- X-rays: binaries -- galaxies: active 
\end{keywords}

\section{Introduction}

\noindent
The study of warped discs dates back to Laplace's (1805) study of the
motions of the satellites of Jupiter, in which he showed that each
satellite precessed around an axis on which the orbit-averaged torques from the
quadrupole moment of the planet and the tidal field from the Sun
cancelled. The locus of the circular rings defined by these axes, now
called the Laplace surface, is the expected shape of a dissipative
low-viscosity disc in this potential \citep[for a review see][]{tre09}.

More recent studies of warped accretion discs began with \cite{bp75},
who pointed out that an accretion disc orbiting a spinning black hole
(BH) would be subject to Lense--Thirring torque if its
orbital axis were not aligned with the spin axis of the BH; this
torque leads to precession of the axis of a test particle on a
circular orbit of radius $r$ at an angular speed
$\bomega=2G\bfL_\bullet/(r^3c^2)$, where $\bfL_\bullet$ is the
angular momentum of the BH\footnote{The quadrupole moment of the BH
  also leads to precession, but this is usually less important as its
  effects fall off faster with radius by a factor $r^{-1/2}$.}.

We call discs `quadrupole' or `Lense--Thirring' discs depending on
which determines the torque from the central body. There are
fundamental differences in the behavior of warped quadrupole and
Lense--Thirring discs. The first is that if the spin axis of the
central body is reversed, the Lense--Thirring torque is also reversed
(eq.\ \ref{eq:lt}) but the quadrupole torque is not (eq.\
\ref{eq:quad}). A second and more fundamental difference is the sign
of the torque: for small inclinations the quadrupole torque induces
retrograde precession of the angular momentum of the disc around the
spin axis of the central body, whereas the Lense--Thirring torque
induces prograde precession. The shape of a steady-state warped disc
is determined by the requirement that the sum of the torques from all
external sources equals the divergence of the angular-momentum
currents from transport within the disc (eqs.\
\ref{eq:ogone}--\ref{eq:ogfour}); thus the difference in sign of the
quadrupole and Lense--Thirring torque leads to fundamental differences
in the geometry of the corresponding discs (\S\ref{sec:invisc}).

Warps are also categorized as `small-amplitude' or
`large-amplitude' depending on whether the amplitude of the warp is
smaller or larger than the disc thickness. The first self-consistent
equations governing warps in viscous fluid discs were derived by
\cite{pp83} in the small-amplitude approximation; their treatment
assumed (as we do in this paper) that the equation of state is barotropic, that
the disc material at radius $r$ is azimuthally symmetric about some
symmetry axis $\bfn(r)$ parallel to the local angular-momentum vector,
that the disc is thin ($H/r\ll1$), and that the time evolution of the
disc is slow ($\p/\p t\ll\Omega$ where $\Omega^2=GM/r^3$ is the
squared angular speed of a Keplerian ring). Among other results
\cite{pp83} found that the behavior of near-Keplerian discs is
complicated by a global resonance between the azimuthal and
radial frequencies $\Omega$ and $\kappa$ of test particles in a
Keplerian potential. Non-resonant behavior requires that
\begin{equation}
\alpha\quad\mbox{or}\quad \left|1-\kappa^2/\Omega^2\right|\ga H/r
\label{eq:nonres}
\end{equation}
where $\alpha$ is the dimensionless Shakura--Sunyaev (1973) viscosity
parameter (eq.\ \ref{eq:alpha}). Most astrophysical discs are
non-resonant in this sense, and we shall assume that this is so in our
analysis. An additional complication, which we shall ignore, is that
the strong, oscillating, shearing flows generated by this
near-resonance are likely to be unstable to the development of
turbulence \citep[see][and references therein]{ol13b}, especially for the low viscosities and large warps that
occupy much of our discussion. 

The equations governing the viscous evolution of thin discs with
large-amplitude warps were derived by \cite{pri92} and
\cite{o99}; a simplified local derivation of the equations is given by
\cite{ol13a}. These authors point out that the evolution of a twisted
disc depends on three conceptually distinct transport coefficients: $\nu_1$ is
the usual viscosity associated with flat accretion discs, which
produces a torque parallel to the local disc normal\footnote{\label{footnote1}Note that
  $\bfn(r)$ is the normal to the orbital plane of the ring at radius
  $r$ but not the normal to the disc surface at radius $r$, which in
  general depends on azimuth.} $\bfn(r)$ that
tends to bring adjacent rings to the same angular speed; $\nu_2$ is
associated with the shear normal to the disc and produces a torque
proportional to $\upartial\bfn/\upartial r$ that tends to bring
adjacent rings to the same orientation; and $\nu_3$ produces a torque
that is proportional to $\bfn\cross\p\bfn/\p r$ and advects angular
momentum in a warped disc. In general these
three transport coefficients are not equal, and a specific model for the stress
tensor in the disc fluid is required to determine their values.
\cite{o99} carries out this determination for Shakura--Sunyaev discs,
in which the shear and bulk viscosity are given by equation
(\ref{eq:alpha}); see for example Fig.\ \ref{fig:two}. However, it is
unclear how directly this treatment applies to real discs,
where the stress tensor is thought to be determined by
magnetohydrodynamic (MHD) turbulence (see  \S\ref{sec:visc}).

The evolution and steady-state shape of warped accretion discs can be
determined by a variety of competing effects: the quadrupole or
Lense--Thirring torque from the central body; mass and angular-momentum
transport through the disc due to viscosity; the tidal field from a
companion object (the Sun for planetary satellites or a stellar
companion for X-ray binary stars); the self-gravity of the disc;
radiation pressure from the central object; magnetic fields; etc.
We shall not consider radiation pressure \citep{pri96} or
magnetic fields \citep{lai99} in this paper, although some of the
phenomena that we describe have analogs when these effects are
important. We distinguish `high-viscosity' from `low-viscosity'
discs depending on whether the torque associated with viscous
angular-momentum transport plays a dominant role in determining the
shape of the warped disc (see \S\ref{sec:approx}).

What we mean by the self-gravity of the disc needs to be
amplified. There are different ways in which discs can be
`self-gravitating'. (i) The radial gravitational force from the disc
can be comparable to the gravity from the host BH, which requires that
the surface density $\Sigma \ga M/r^2$; this case is not relevant for most accretion
discs and we shall not discuss it further. (ii) Within the disc, the
vertical gravitational force from the disc can be comparable to the
vertical gravity from the BH; this requires that the density in the
disc is of order $M/r^3$ or that Toomre's (1964) $Q$ parameter (eq.\
\ref{eq:toomredef}) is of order unity. Models of accretion discs with
$Q\simeq 1$ were first described by \cite{pac78}; in accretion discs
surrounding supermassive BHs in active galactic nuclei (AGN) this
condition is likely to be satisfied at distances exceeding $\sim
0.01\,\mbox{pc}$, and such discs may fragment into stars
\citep{2003MNRAS.339..937G}.  (iii) The apsidal and/or nodal
precession rate of the disc may be dominated by self-gravity; for AGN
accretion discs this requires far less mass than cases (i) or (ii) and
this is the case that we focus on here. 

Remarkably, almost all previous studies of warped Lense--Thirring
discs follow Bardeen and Petterson in considering only torques from
the central body and viscous torques in their analyses. We shall show
that the other two effects listed in the preceding
paragraph -- gravitational torques from the companion and the
self-gravity of the disc -- can introduce qualitatively new phenomena
in the behavior of warped discs surrounding stellar-mass and
supermassive BHs, respectively. In particular, (i) warped
low-viscosity discs exhibit a sharp depression in their surface
density near the radius where the warp is strongest; (ii)
steady-state Lense--Thirring discs do not exist, at least within the
standard thin disc description, for viscosities below a critical value
that depends on the obliquity (the angle between the BH spin angular
momentum and the companion orbital angular momentum); (iii) warped low-viscosity discs in
which self-gravity is important can develop strong short-wavelength
bending waves.

As a preliminary step, \S\S\,\ref{sec:ext} and \ref{sec:invisc}
derive the steady-state properties of warped discs in which viscosity
is negligible. Then \S\,\ref{sec:approx} provides a broad-brush
overview of the competing effects that determine the behavior of
warped discs. Section \ref{sec:visc} derives the equations of motion
for a thin, viscous disc subjected to external torques, following
\cite{pri92} and \cite{o99}, and \S\,\ref{sec:results} describes our
numerical methods and the results for both quadrupole and
Lense--Thirring discs in systems with a binary companion. Section
\ref{sec:sg} describes the behavior of self-gravitating warped
discs. Section \ref{sec:other} relates our findings to earlier work on
warped accretion discs.  Sections \ref{sec:xrb} and \ref{sec:agn}
apply our results to accretion discs around stellar-mass BHs in binary
systems, around supermassive black holes in AGN.  Finally,
\S\ref{sec:summary} contains a brief summary of our conclusions.

\subsection{External torques}

\label{sec:ext}

\noindent
In this paper we consider three types of external torque that can warp
an accretion disc. In each case we shall assume that the torque is
weak -- the fractional change per orbit in the angular momentum of an
orbiting fluid element is small -- so we can work with the
orbit-averaged torque. In particular we define $\bfT(r,\bfn,t)$ to be
the torque per unit mass averaged over a circular orbit at radius $r$
with orbit normal $\bfn$.

\paragraph*{Quadrupole torque:} In the system examined
by Laplace, the central body is a planet of mass $M$, radius $R_p$,
and quadrupole gravitational harmonic $J_2$.  If the planet's spin
axis is along $\bfn_p$, the torque per unit mass on an orbiting test particle is
\begin{equation}
\bfT_p=\frac{\epsilon_p}{r^3}(\bfn\cdot\bfn_p)\,\bfn\cross\bfn_p
\quad\mbox{where}\quad
\epsilon_p=\frac{3}{2}GMJ_2R_p^2\,.
\label{eq:quad}
\end{equation}
The quadrupole torque is also relevant to circumbinary accretion
discs; in the case of a binary with masses $M_1$ and $M_2$ on a
circular orbit with separation $a\ll r$, we replace $M$ by $M_1+M_2$
and $J_2R_p^2$ by $\frac{1}{2}M_1M_2\,a^2/(M_1+M_2)^2$. 

\paragraph*{Lense--Thirring torque:} The central body can also be a BH
of mass $M$ and angular momentum $\bfL_\bullet=GM^2a_\bullet\,\bfn_\bullet/c$ where
$c$ is the speed of light, $\bfn_\bullet$ is the spin axis of the BH
and $0\le a_\bullet<1$ is the dimensionless spin parameter of the BH. 
The angular momentum of a test particle orbiting the BH
precesses as if it were subject to a classical torque (the
Lense--Thirring torque; see \citealt{llfields})
\begin{equation}
\bfT_{\rm LT}=-\frac{\epsilon_{\rm LT}}{r^{5/2}}\,\bfn\cross\bfn_\bullet
\quad\mbox{where}\quad \epsilon_{\rm LT}={2(GM)^{5/2}a_\bullet\over
  c^3}=2R_g^{5/2}c^2a_\bullet\,,
\label{eq:lt}
\end{equation}
where $R_g\equiv GM/c^2\ll r$ is the gravitational radius of the
BH. 

\paragraph*{Companion torque} The central body, whether a planet or a
BH, may be accompanied by a companion star of mass $\mstar$, on a
circular orbit with radius $\rstar\gg r$. Then the gravitational
potential of the companion can be approximated by its quadrupole
component, which after averaging over the companion orbit yields a
torque
\begin{equation}
\bfT_\star=\epsilon_\star r^2
(\bfn\cdot\bfn_\star)\,\bfn\cross\bfn_\star \quad\mbox{where}\quad
\epsilon_\star=\frac{3G\mstar}{4\rstar^3}.
\label{eq:comp}
\end{equation}

\subsection{Inviscid discs} 

\label{sec:invisc}

\noindent
Following Laplace, we first consider a thin disc of material orbiting
a planet with non-zero obliquity (the obliquity is
$\cos^{-1}\bfn_p\cdot\bfn_\star$).  The disc is subject to torques
from the quadrupole moment of the planet, $\bfT_p$ (eq.\
\ref{eq:quad}), and from the companion star around which the planet
orbits, $\bfT_\star$ (eq.\ \ref{eq:comp}). In the absence of pressure,
viscosity, self-gravity, or other collective effects in the disc, the
fluid rings at different radii precess independently, so the disc
cannot retain its coherence unless the total torque $\bfT_\star +
\bfT_p=0$ at each radius. This requires
\begin{equation}
r^5 (\bfn\cdot\bfn_\star)\, \bfn\cross\bfn_\star +
\frac{\epsilon_p}{\epsilon_\star}(\bfn\cdot\bfn_p)\,
\bfn\cross\bfn_p=0,
\end{equation}
which can be rewritten as
\begin{equation}
\left(\frac{r}{r_w}\right)^5 (\bfn\cdot\bfn_\star)\,
\bfn\cross\bfn_\star + (\bfn\cdot\bfn_p)\,\bfn\cross\bfn_p=0 \quad\mbox{where} \quad r_w^5\equiv\frac{\epsilon_p}{\epsilon_\star}=2
J_2\frac{M}{\mstar}R_p^2\rstar^3
\label{eq:lap}
\end{equation}
defines the characteristic radius $r_w$ at which the warp is most prominent \citep{pmg66}. 

We restrict ourselves to the usual case in which the disc normal $\bfn(r)$ is coplanar
with $\bfn_p$ and $\bfn_\star$ (for a more general discussion see \citealt{tre09}).
Then the unit vectors $\bfn(r)$, $\bfn_p$, $\bfn_\star$ can be
specified by their azimuthal angles in this plane, $\phi(r)$,
$\phi_p$, $\phi_\star$. Without loss of generality we may assume
$\phi_\star=\half\upi$, so the obliquity is $\phi_p-\phi_\star=\phi_p-\half\upi$.
Then equation (\ref{eq:lap}) can be rewritten as
\begin{equation}
\left(\frac{r}{r_w}\right)^5 = \frac{\sin 2(\phi-\phi_p)}{\sin
  2\phi}.
\label{eq:lapa}
\end{equation} 

\begin{figure}
\includegraphics[width=0.5\textwidth]{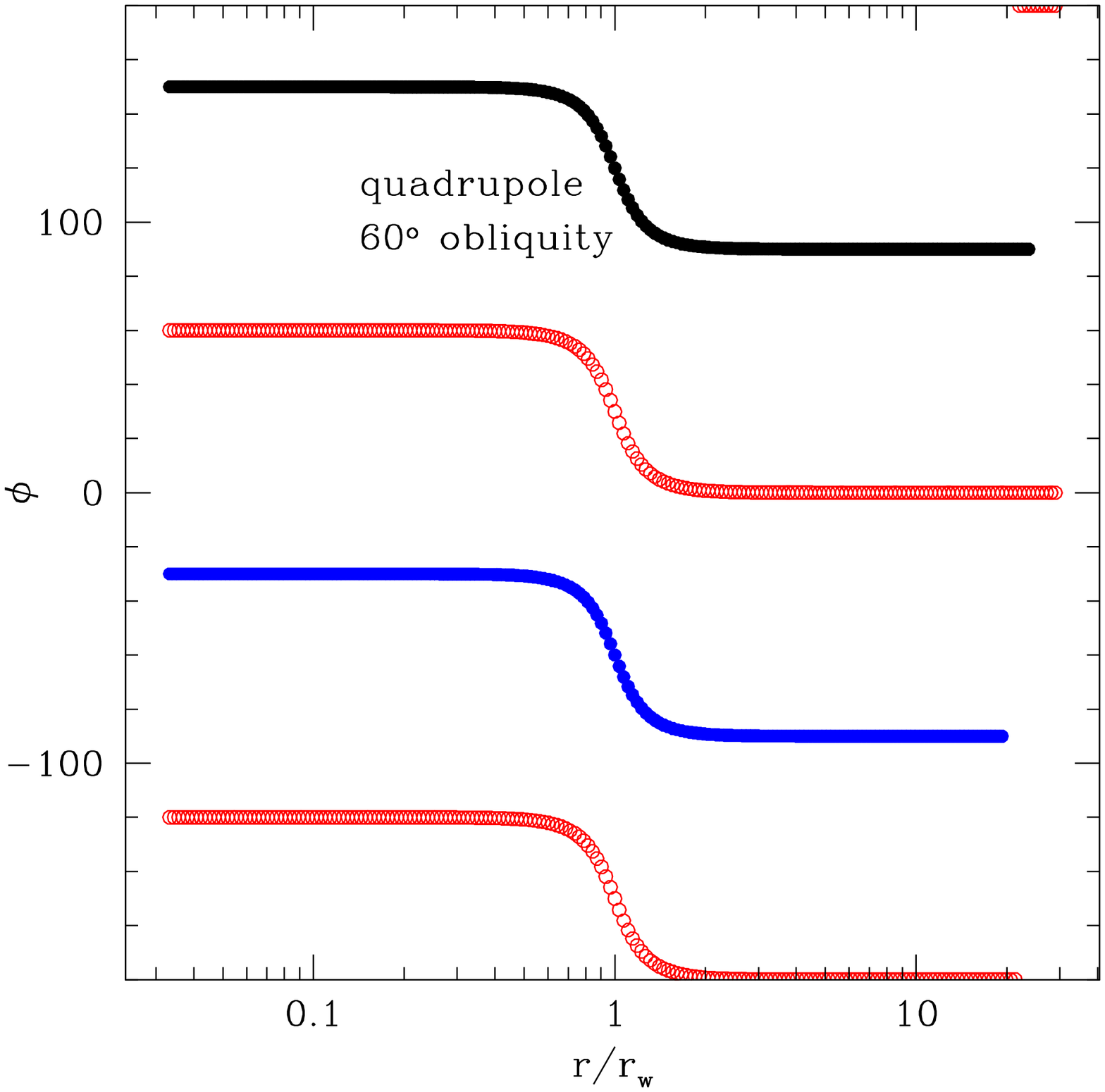}
\includegraphics[width=0.5\textwidth]{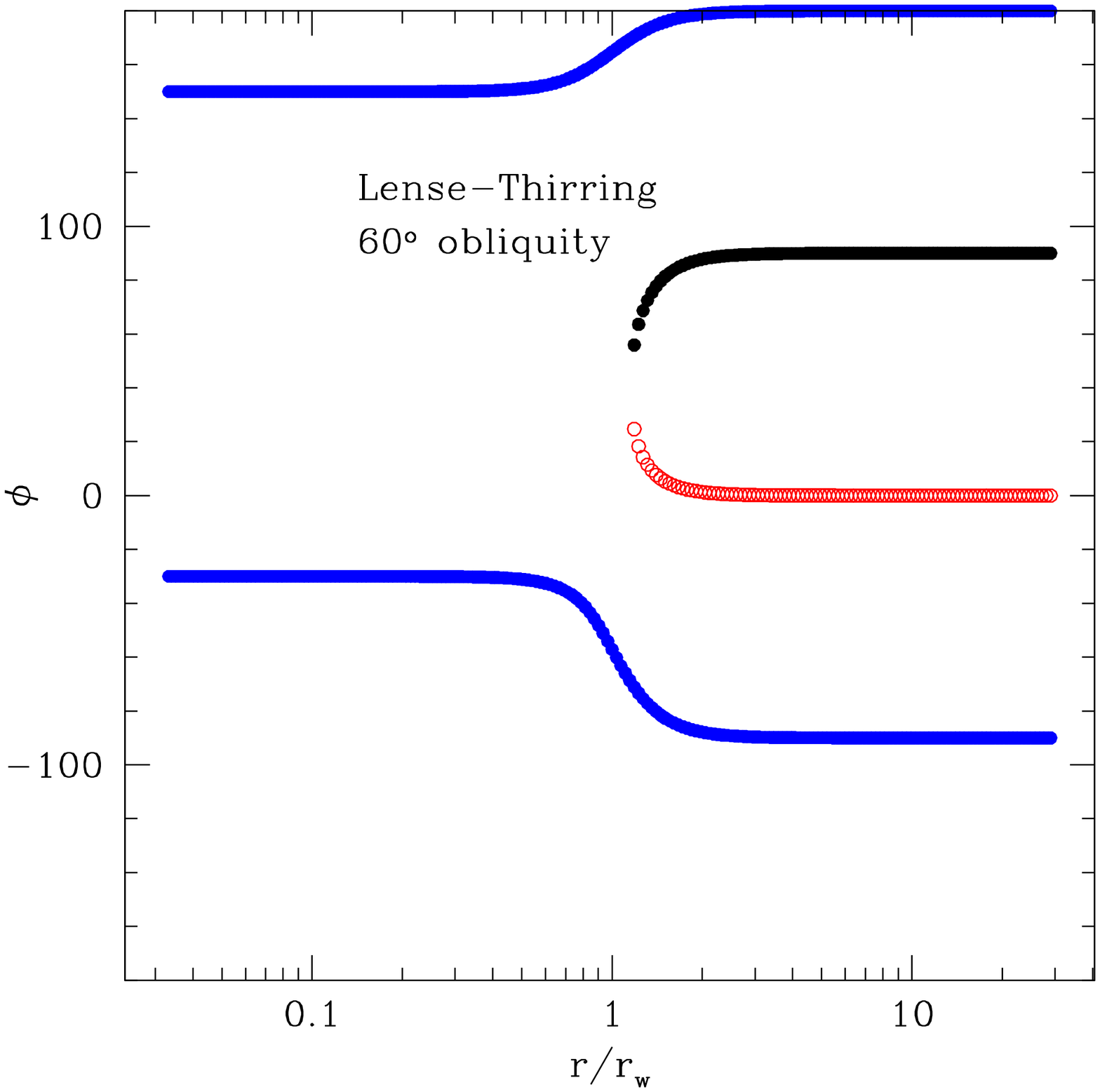}
\caption{(left) The orientation of a stationary, inviscid disc
  orbiting a planet that has an obliquity of $60^\circ$. An orbit with
  angular momentum aligned with the planetary orbit has azimuthal
  angle $\phi=90^\circ$ and an orbit aligned with the planetary
  equator has $\phi=90^\circ+60^\circ=150^\circ$. The black solid
  circles denote the classical Laplace surface, the blue circles
  denote the same spatial surface as traced by retrograde orbits,
  and the red open circles denote dynamically unstable
  surfaces. (right) The same as the left panel, but for an inviscid
  disc orbiting a spinning BH; like the planet, the BH
  orbits a companion star with an obliquity of $60^\circ$.}
\label{fig:one}
\end{figure} 

The solutions to equation (\ref{eq:lapa}) are shown in the left panel
of Fig.\  \ref{fig:one} for obliquity $\phi_p-\phi_\star=60^\circ$.
The `classical' Laplace surface, shown as solid black circles, is
aligned with the planet's orbit around the star at large radii
($\phi\to\half\upi$ as $r\to\infty$). The surface shown by solid blue
circles is similar, but composed of retrograde orbits (the disc
angular-momentum vector is anti-aligned with the planetary orbital
angular momentum at large radii, and anti-aligned with the planetary
spin at small radii). The surfaces shown by open red circles are also
solutions of equation (\ref{eq:lapa}) but they are unstable to small
perturbations in $\bfn$ \citep{tre09}, and we will not consider them
further. On the classical Laplace surface, the azimuth of the disc
normal $\phi$ increases smoothly and continuously from $\phi_\star$ to
$\phi_p$, so that the disc plane gradually twists from the orbital
plane of the planet to the equatorial plane of the planet as its
radius shrinks.

We next carry out the analogous derivation for an inviscid thin disc
orbiting a spinning BH with a companion star. The disc is subject to
Lense--Thirring torque, $\bfT_{\rm LT}$ (eq.\ \ref{eq:lt}), and torque
from the companion star, $\bfT_\star$ (eq.\ \ref{eq:comp}). The
equilibrium shape defined by $\bfT_\star+\bfT_{\rm LT}=0$ is given by
\begin{equation}
r^{9/2} (\bfn\cdot\bfn_\star)\, \bfn\cross\bfn_\star -
\frac{\epsilon_{\rm LT}}{\epsilon_\star}\bfn\cross\bfn_\bullet=0
\end{equation}
which can be rewritten as 
\begin{equation}
\left(\frac{r}{r_w}\right)^{9/2} (\bfn\cdot\bfn_\star)\,
\bfn\cross\bfn_\star - \bfn\cross\bfn_\bullet=0 \quad\mbox{where}
\quad r_w^{9/2}=\frac{\epsilon_{\rm LT}}{\epsilon_\star}=\frac{8a_\bullet}{3}\frac{M}{\mstar}R_g^{3/2}\rstar^3.
\label{eq:kerr}
\end{equation}
The analog to equation (\ref{eq:lapa}) is
\begin{equation}
\left(\frac{r}{r_w}\right)^{9/2}=-\frac{2\sin(\phi-\phi_\bullet)}{\sin2\phi},
\label{eq:lapk}
\end{equation}
where $\phi_\bullet$ is the azimuthal angle of the BH spin
axis. The obliquity is $\phi_\bullet-\phi_\star=\phi_\bullet-\half\upi$.

The solutions to equation (\ref{eq:lapk}) are shown in the right panel
of Fig.\  \ref{fig:one} for obliquity
$\phi_\bullet-\phi_\star=60^\circ$. In contrast to the quadrupole
case, the solution that is aligned with the companion-star orbit at
large radii ($\phi\to\half\upi$ as $r\to\infty$, shown as black filled
circles) terminates just outside the characteristic radius $r_w$ (this
solution is mirrored by an unstable solution, shown by open red
circles, that has no relevance to our discussion). The solution that
is aligned with the equator of the BH at small radii, shown as the
upper set of filled blue circles, approaches $\phi=\upi$ at large
radii; in other words the disc is perpendicular to the companion-star
orbital plane, which is inconsistent with the expectation that the
disc is fed by material lost from the companion.  Material spiraling
in from the companion star along the black sequence of points in the
right panel of Fig.\  \ref{fig:one} must therefore jump to one of the
two blue sequences before proceeding inwards to the
BH\footnote{J. Touma (private communication) points out that the time
  evolution of the orbit normals in the Lense--Thirring disc is the
  same as that of Colombo's top, which describes the behavior of the
  spin axis of the Moon due to the torque from the Earth on the lunar
  figure and precession of the lunar orbit due to the Sun
  \citep{col66,hen87}. The solutions shown in the right panel of
  Fig.\  \ref{fig:one} correspond to the Cassini states of the Moon,
  of which there are two or four depending on whether the lunar
  semimajor axis is less than or greater than 34 Earth radii
  \citep{ward}.}.

The lower blue sequence represents a solution in which
the disc angular momentum is anti-aligned with the BH spin at small
radii ($\phi=\phi_\bullet-\upi$) and anti-aligned with the orbital
angular momentum of the companion at large radii. This is equivalent
to a solution in which the obliquity is $120^\circ$ and the disc
angular momentum is aligned with the BH spin at small radii and the
companion's orbital angular momentum at large radii. Thus a smooth
surface similar to the classical Laplace surface seen in the left
panel of Fig.\  \ref{fig:one} exists around a spinning BH if and only if the obliquity
exceeds $90^\circ$.

These conclusions raise two obvious questions: how is this unusual behavior
related to the standard Bardeen--Petterson analysis of a warped
accretion disc orbiting a spinning BH? And how do warped accretion discs
actually behave in real astrophysical systems?

\subsection{An approximate analysis of viscous warped discs}

\label{sec:approx}

\noindent
To show the relation between the findings of the preceding subsection
and the Bardeen--Petterson treatment of viscous warped discs, we
examine the approximate strength of the torques from various sources.

Suppose that the disc is strongly warped near some radius
$r$. The torque per unit mass due to a companion is
(eq.\ \ref{eq:comp})
\begin{equation}
T_\star\simeq \frac{G\mstar r^2}{\rstar^3},
\end{equation}
where we have neglected all factors of order unity. Similarly, the torque from
the quadrupole moment of the central body is (eq.\ \ref{eq:quad})
\begin{equation}
T_p\simeq \frac{GMJ_2R_p^2}{r^3};
\end{equation}
and the Lense--Thirring torque is (eq.\ \ref{eq:lt})
\begin{equation}
T_{\rm LT}\simeq \frac{R_g^{5/2}c^2a_\bullet}{r^{5/2}}.
\label{eq:fffggg}
\end{equation}
The torque per unit mass due to viscous stress is $T_v\simeq
\eta\Omega/\rho$ where $\eta$ is the viscosity and $\rho$ is the
density in the disc. In the Shakura--Sunyaev $\alpha$-model of
viscosity (eq.\ \ref{eq:alpha}) $\eta=\alpha \rho c_s^2$ where $c_s$
is the sound speed, and $\alpha$ is a constant, typically assumed to
be $\sim 0.1$. However, the Shakura--Sunyaev model was developed to
model viscous forces in the disc arising from Keplerian shear, whereas
the warp shape is determined by viscous forces due to much smaller
shears normal to the disc plane. To represent the second kind of force
we use an $\alpha$-model with a different parameter $\alpha_\perp$
(for small-amplitude warps $\alpha_\perp=\frac{1}{2}\alpha^{-1}$; see
eq.\ \ref{eq:qalpha}). Thus
\begin{equation}
T_v\simeq \alpha_\perp c_s^2.
\label{eq:visct}
\end{equation}
For simplicity we shall usually assume that the disc is isothermal, in
which case the viscous torque is independent of radius.  Finally, the torque
per unit mass due to the self-gravity of the disc is roughly
\begin{equation}
T_{sg} \simeq \upi G\Sigma r.
\end{equation}
where $\Sigma$ is the surface density near radius $r$.

\paragraph*{Viscous quadrupole discs with a companion} 
The quadrupole torque $T_p$ decreases with radius, while
the torque from the companion $T_\star$ increases with radius. The two
are equal at
\begin{equation}
r_w\simeq \left(J_2\frac{M}{\mstar}R_p^2\,\rstar^3\right)^{1/5}.
\end{equation}
which agrees with the precise definition of the warp radius in equation
(\ref{eq:lap}) to within a factor of order unity. Since the viscous
torque $T_v$ is independent of radius in an isothermal disc, and one of
$T_\star$, $T_p$ is always larger than $T_\star(r_w)$, the viscous
torque is always smaller than the torque due to the central body or
the companion if $\beta\alpha_\perp <1$, where 
\begin{equation}
\beta\equiv
\frac{T_v/\alpha_\perp}{T_\star(r_w)}=\frac{c_s^2R_p}{GMJ_2^{2/5}}
\left(\frac{\rstar}{R_p}\right)^{9/5}\left(\frac{M}{\mstar}\right)^{3/5}. 
\label{eq:betaone}
\end{equation}
This agrees with the precise definition of $\beta$ that we give later
in the paper (eq.\ \ref{eq:qdef}) to within 1 per cent. In
the terminology introduced at the start of the paper, a disc with
$\beta\alpha_\perp\la 1$ is a `low-viscosity' disc.

\paragraph*{Viscous Lense--Thirring discs with a companion} The
Lense--Thirring torque $T_{\rm LT}$ and the companion torque $T_\star$
are equal at
\begin{equation}
r_w\simeq \left(a_\bullet\frac{M}{\mstar} R_g^{3/2}\rstar^3\right)^{2/9},
\end{equation}
and the ratio of the viscous torque to the Lense--Thirring or companion torque at
$r_w$ is then $\beta\alpha_\perp$ where 
\begin{equation}
\beta\equiv \frac{T_v/\alpha_\perp}{T_\star(r_w)}=
\frac{c_s^2}{c^2a_\bullet^{4/9}}\left(\frac{\rstar}{R_g}\right)^{5/3}\left(\frac{M}{\mstar}\right)^{5/9},
\label{eq:betatwo}
\end{equation}
consistent with the precise definition in equation (\ref{eq:qdef}) to
within 15 per cent.

We expect that the shape of a low-viscosity disc
($\beta\alpha_\perp\la 1$) is determined by the competition
between the torque from the central body (quadrupole or
Lense--Thirring torque) and the torque from the companion, rather than
by viscous torques. On the other hand the surface-density distribution
in a warped disc is always determined by the viscous torque, no matter
how small, since the other two torques both scale linearly with the
surface density and hence do not establish the surface-density
distribution.

The usual Bardeen--Petterson description implicitly assumes that
$\beta\alpha_\perp\gg1$ and neglects the companion torque. In this case the warp
will be strongest at a smaller radius $r_w'$ given by
\begin{equation}
r_w'\simeq\left\{\begin{array}{ll} r_w/(\alpha_\perp\beta)^{1/3}\simeq
    (J_2R_p^2GM/\alpha_\perp c_s^2)^{1/3};
    & \qquad\mbox{quadrupole disc} \\[10pt]
r_w/(\alpha_\perp\beta)^{2/5}\simeq (a_\bullet/\alpha_\perp)^{2/5}\left(c/c_s\right)^{4/5}R_g
      & \qquad \mbox{Lense--Thirring disc}.
\end{array}\right.
\label{eq:rwprime}
\end{equation}

\paragraph*{Viscous Lense--Thirring discs with self-gravity}
In accretion discs surrounding supermassive BHs at the centres of
galaxies, there is no companion body (except in the case of a binary
BH; see \S\ref{sec:bbh}) . Thus the torque $T_\star$ can be neglected. However, the disc can
be massive enough that its self-gravity plays a role in determining
its shape.  In plausible disc models the surface density falls off
slowly enough that this torque increases outward (see
\S\ref{sec:agn}), and equals the Lense--Thirring torque at
\begin{equation}
r_w\simeq \left[\frac{a_\bullet R_g^{5/2}c^2}{\upi G\Sigma(r_w)}\right]^{2/7};
\label{eq:rwself}
\end{equation}
note that this is an implicit equation for the warp radius $r_w$ since
the surface density depends on radius.  The ratio of the viscous torque, equation
(\ref{eq:visct}), to the Lense--Thirring and self-gravity torques at
$r_w$ is then $\gamma\alpha_\perp$, where 
\begin{equation}
  \gamma\equiv
  \frac{T_v/\alpha_\perp}{T_{sg}(r_w)}=\frac{c_s^2}{\upi G\Sigma r}\bigg|_{r_w}.
\label{eq:betaself}
\end{equation}
Note that $\gamma\simeq Q (H/r)$ where $Q$ is Toomre's parameter (eq.\
\ref{eq:toomredef}) and $H=c_s/\Omega$ is the disc thickness. Thus the
viscosity becomes low (in the sense that $\gamma\ll1$) in thin discs ($H/r\ll1$)
long before they become gravitationally unstable ($Q<1$).

\section{Evolution of viscous discs with companions}

\subsection{Evolution equations}

\label{sec:visc}

\noindent
The equations that describe the evolution of a warped, thin accretion
disc are derived by \cite{pri92}, \cite{o99}, and \cite{ol13a}. Our starting point
is \cite{o99}'s equations (121) and (122). The first of these is the
equation of continuity
\begin{equation}
2\upi r{\p\Sigma\over\p t}+{\p C_M\over\p r}=0, \qquad
C_M\equiv 2\upi r\Sigma v_r,
\label{eq:ogone}
\end{equation}
where $\Sigma(r,t)$ is the surface density, $v_r(r,t)$ is the radial
drift velocity, and $C_M(r,t)$ is the mass current (rate of outward
flow of disc mass through radius $r$). The second is an equation for
angular momentum conservation,
\begin{equation}
2\upi r{\p\bfL\over \p t} +\frac{\p\bfC_L}{\p r}=2\upi r\Sigma\bfT,
\label{eq:ogtwo}
\end{equation}
where $\Omega(r)\equiv (GM/r^3)^{1/2}$ is the Keplerian angular
speed, $\bfL=\Sigma r^2\Omega\,\bfn$ is the angular momentum per unit
area, $\bfT$ is the torque per unit mass from sources external to the
disc, and $\bfC_L$ is the angular-momentum current, given by the sum
of advective and viscous currents, 
\begin{align}
\bfC_L\equiv &\bfC_{\rm adv}+\bfC_{\rm visc}, \nonumber \\[10pt]
\bfC_{\rm adv}(r,t)= & 2\upi r^3\Omega\Sigma v_r\,\bfn = r^2\Omega\,\bfn\,C_M,\nonumber \\
\bfC_{\rm visc}(r,t)= & -2\upi r^2\Sigma c_s^2 \Big(Q_1\bfn +
      Q_2r{\p\bfn\over\p r} + Q_3r\, \bfn\cross{\p\bfn\over\p
        r}\Big).
\label{eq:ogfour}
\end{align}
Here $c_s$ is the sound speed, which is constant in an isothermal disc
(as we shall assume from now on), and as usual\footnotemark[2] $\bfn(r,t)$ is the unit vector normal to
the disc at radius $r$. The dimensionless coefficients $Q_1$, $Q_2$,
$Q_3$ depend on the equation of state, the viscosity, and the
warp $\psi\equiv r|\p\bfn/\p r|$. For a flat Keplerian disc, $Q_1$ is related to the kinematic
viscosity by $\nu=-\frac{2}{3}Q_1c_s^2/\Omega$ and the mean-square
height of the disc above the midplane is $H^2=c_s^2/\Omega^2$. 

These equations are based on the assumptions \citep{o99} that (i) the
disc is thin, $H/r\ll1$; (ii) the fluid obeys the compressible
Navier--Stokes equation; (iii) the fluid equation of state is
barotropic, i.e., the viscosity is dynamically important but not
thermodynamically important; (iv) the disc is non-resonant in the
sense of equation (\ref{eq:nonres}). In the calculations below we shall also
assume that (v) the viscosity is described by the Shakura--Sunyaev
$\alpha$-model, that is, the shear and bulk viscosities $\eta$ and
$\zeta$ are related to the pressure $p$ by
\begin{equation}
  \eta=\alpha\,p/\Omega, \quad \zeta=\alpha_b\, p/\Omega,
\label{eq:alpha}
\end{equation}
where $\alpha$ and $\alpha_b$ are constants. For a flat, isothermal
disc the kinematic viscosity is $\nu=\eta/\rho=\alpha c_s^2/\Omega$,
so $\alpha=-\frac{2}{3}Q_1$. 

Now take the scalar product of (\ref{eq:ogtwo}) with $\bfn$. Since
$\bfn\cdot\bfn=1$, $\bfn\cdot \p\bfn/\p t=\bfn\cdot \p\bfn/\p r=0$.
Moreover $\bfn\cdot\bfT=0$ for the Lense--Thirring torque and for any
torque arising from a gravitational potential, so we
shall assume that this condition holds in general. We also use
equation (\ref{eq:ogone}) to eliminate $\p\Sigma/\p t$. The result is
an expression for the mass current,
\begin{equation}
C_M = 2\upi r\Sigma
v_r=-\frac{2}{r\Omega}\bfn\cdot\frac{\p\bfC_{\rm visc}}{\p
  r}=\frac{4\upi c_s^2}{r\Omega}\frac{\p}{\p
  r}\left(\Sigma r^2 Q_1\right)-\frac{4\upi\Sigma c_s^2r^2}{\Omega}
  Q_2\bigg|\frac{\p\bfn}{\p r}\bigg|^2.
\label{eq:masscurr}
\end{equation}

We now introduce several new variables: the dimensionless radius
$x\equiv r/r_w$ with the warp radius $r_w$ given by (\ref{eq:lap}) or (\ref{eq:kerr});
the dimensionless time $\tau\equiv t\, c_s^2/(GM r_w)^{1/2}$
(roughly, for a Shakura--Sunyaev disc with $\alpha\sim 1$ this is time
measured in units of the viscous diffusion time at the warp radius);
and $y(r,t)\equiv \Sigma(r,t)(GM r_w)^{1/2}$ (with dimensions
of angular momentum per unit area). Equation (\ref{eq:ogtwo}) becomes
\begin{equation}
\frac{\p\bfL}{\p\tau}+\frac{1}{x}\frac{\p}{\p x}(\bfc_{\rm
  visc}+x^{1/2}c_M\bfn)=\frac{y}{\beta}\left\{\begin{array}{ll}x^2(\bfn\cdot\bfn_\star)\bfn\cross\bfn_\star+x^{-3}(\bfn\cdot\bfn_p)\bfn\cross\bfn_p 
& \quad \mbox{quadrupole}  \\[5pt]
x^2(\bfn\cdot\bfn_\star)\bfn\cross\bfn_\star-x^{-5/2}\bfn\cross\bfn_\bullet
& \quad \mbox{Lense--Thirring} 
\end{array}\right.
\label{eq:ogthree}
\end{equation}
where $\bfn=\bfL/|\bfL|=\bfL/(\Sigma r^2\Omega)=\bfL/(yx^{1/2})$, $y=|\bfL|/x^{1/2}$,
\begin{align}
\bfc_{\rm visc}\equiv &\;
\frac{1}{2\upi c_s^2}\bigg(\frac{GM}{r_w^3}\bigg)^{1/2} \bfC_{\rm visc}=-yx^2\bigg(Q_1\bfn+Q_2 x\frac{\p\bfn}{\p x}
+Q_3x\bfn\cross\frac{\p \bfn}{\p x}\bigg), \nonumber \\[10pt]
c_M\equiv & \;\frac{GM}{2\upi r_wc_s^2}C_M=-2x^{1/2}\bfn\cdot\frac{\p\bfc_{\rm visc}}{\p x}=2x^{1/2}\bigg[\frac{\p}{\p
  x}\left(y x^2Q_1\right)-y x^3Q_2\Big|\frac{\p\bfn}{\p x}\Big|^2\bigg].
\label{eq:cldef}
\end{align}

 The dimensionless parameter $\beta$ is given by  
\begin{equation}
\beta\equiv
 \frac{4M}{3\mstar}\frac{c_s^2r_w}{GM}\left(\frac{\rstar}{r_w}\right)^3=\left\{\begin{array}{ll}
\displaystyle
\frac{2^{8/5}}{3J_2^{2/5}}\frac{c_s^2R_p}{GM}\left(\frac{\rstar}{R_p}\right)^{9/5}\left(\frac{M}{\mstar}\right)^{3/5}
  &\qquad\mbox{quadrupole} \\[15pt]\displaystyle
\frac{2^{2/3}}{3^{5/9}a_\bullet^{4/9}}\frac{c_s^2}{c^2}\left(\frac{\rstar}{R_g}\right)^{5/3}\left(\frac{M}{\mstar}\right)^{5/9}
&\qquad\mbox{Lense--Thirring}
\end{array}\right.
\label{eq:qdef}
\end{equation}
and represents the ratio of the strength of the viscous torque to the
external torque at the characteristic warp radius $r_w$ (cf.\ eqs.\
\ref{eq:betaone} and \ref{eq:betatwo}).

Equation (\ref{eq:ogthree}) is a parabolic partial differential
equation for the three components of $\bfL$.
The dimensionless viscosity coefficients $Q_i$
are functions of the equation of state and of the warp $\psi\equiv x|\upartial\bfn/\upartial x|$
\citep{o99}. Ogilvie shows that for an isothermal $\alpha$-disc and
small warps ($\psi\ll1$), 
\begin{equation}
Q_1=-\frac{3\alpha}{2} + \mbox{O}(\psi^2),\qquad 
Q_2=\frac{1+7\alpha^2}{\alpha(4+\alpha^2)}+\mbox{O}(\psi^2)=
\frac{1}{4\alpha} +\mbox{O}(\alpha,\psi^2).
\label{eq:qsmall}
\end{equation}

We shall also examine a simplified set of equations that appear to
contain most of the important physics of equations
(\ref{eq:ogthree})--(\ref{eq:qdef}). In these equations (i) we examine
only the steady-state disc, that is, we set $\p\bfL/\p t=0$ in
equation (\ref{eq:ogthree}); (ii) we set $Q_3=0$, since it appears to
play no important role in the dynamics; and (iii) we neglect the
dependence of $Q_1$ and $Q_2$ on the warp $\psi$, that is, we treat
them as constants. The steady-state assumption implies that the mass
current $c_M$ is a constant of the problem, independent of radius. We
have
\begin{align}
\frac{dy}{dx}&+y\left(\frac{2}{x}-\frac{Q_2x}{Q_1}|d\bfn/dx|^2\right)=\frac{c_M}{2Q_1x^{5/2}},
\nonumber \\[10pt]
\frac{d^2\bfn}{dx^2} &+\frac{d\bfn}{dx}\left[\frac{Q_1/Q_2+3}{x}
  -\frac{c_M}{Q_2x^{5/2}y}+\frac{d\log y}{dx}\right]+ |d\bfn/dx|^2\bfn
\nonumber \\
&= \qquad -\frac{1}{\beta Q_2}\left\{\begin{array}{ll}(\bfn\cdot\bfn_\star)\bfn\cross\bfn_\star+x^{-5}(\bfn\cdot\bfn_p)\bfn\cross\bfn_p
& \qquad \mbox{quadrupole}  \\[5pt]
(\bfn\cdot\bfn_\star)\bfn\cross\bfn_\star-x^{-9/2}\bfn\cross\bfn_\bullet
& \qquad \mbox{Lense--Thirring} 
\end{array}\right.
\label{eq:simple}
\end{align}
The three components of the unit vector $\bfn$ are related by
the constraint $|\bfn|=1$. 

This simplified model is similar to Pringle's (1992) equations of
motion, in which there are two viscosities $\eta$ and $\eta_\perp$ (in
Pringle's notation, these are $\rho\nu_1$ and $\rho\nu_2$), the first
of which is associated with the Keplerian shear and the second with
shear perpendicular to the disc caused by a warp. In an $\alpha$-disc
model $\eta=\alpha \rho c_s^2$ and $\eta_\perp=\alpha_\perp\rho c_s^2$
and the two models are equivalent if 
\begin{equation}
Q_1=-\frac{3\alpha}{2}, \qquad Q_2=\frac{\alpha_\perp}{2}.
\label{eq:qalpha}
\end{equation}
If $\alpha\ll1$ and the warp is small, equation (\ref{eq:qsmall})
implies that $\alpha_\perp=\frac{1}{2}\alpha^{-1}$ \citep{pp83,o99}.

Although we adopt this formalism, one should keep in mind that
angular-momentum transport in real accretion discs is thought to be
driven by MHD turbulence, which may not be well approximated by an
isotropic viscosity -- or if it is, the viscosity may not be well
approximated by the Shakura--Sunyaev $\alpha$-model.  Some support for
this formalism is provided by local, non-relativistic MHD simulations
that examine the decay of an imposed epicyclic oscillation
\citep{tor00}.  Global, general-relativistic MHD simulations have
tended to show solid-body precession rather than Bardeen--Petterson
alignment, although most of these correspond to the resonant regime
$\alpha < H/r$ (cf.\ eq.\ \ref{eq:nonres}), which we exclude
\citep[e.g.][]{fra07}.  More recently, global but non-relativistic MHD
calculations with an approximate treatment of Lense--Thirring
precession have been performed by Sorathia et al.\ (submitted to ApJ;
see also \citealt{2013ApJ...768..133S}).  They find that diffusive
damping of vertical shear is much less important than the derivation
of the Pringle--Ogilvie equations implies.  This in turn implies that the 
Pringle--Ogilvie + Shakura--Sunyaev formalism overestimates the
strength of viscous torques when $\alpha\ll1$ and so the importance of
tidal torques and self-gravity in accretion discs is even greater than
we find below. 

\subsection{Numerical methods}

\label{sec:methods}

\paragraph*{Steady-state discs} We have solved the simplified ordinary
differential equations (\ref{eq:simple}) for steady-state discs with
constant viscosity coefficients and $Q_3=0$. We find the numerical
solution over a range of dimensionless radii $[x_a,x_b]$; typically we
choose $x_b=1/x_a=30$, although in some cases where the viscosity is
large we cover a larger range to ensure that the disc is not still
warped at either end of the integration range. The viscosity
coefficients $Q_1$ and $Q_2$ are usually fixed at their values for an
unwarped disc with $\alpha=0.2$, $\alpha_b=0$, in which case
$Q_1=-0.3$, $Q_2=1.58416$.  The equations are unchanged under the
rescaling $y(x)\to\lambda y(x)$, $c_M\to \lambda c_M$, so the
normalization of the mass current $c_M$ can be chosen arbitrarily
apart from the sign. We are interested in the case in which mass flows
into the BH, so we set $c_M=-1$.

Seven boundary conditions are required for the one first-order and
three second-order equations.  In the region $x\ll1$ where external
torques are negligible, the disc is assumed to be flat,
$d\bfn/dx=0$. Then the first of equations (\ref{eq:simple}) has the
solution
\begin{equation}
y(x)=\frac{c_M}{Q_1x^{3/2}} + \frac{k}{x^2},
\label{eq:qqwwrr}
\end{equation}
where $k$ is an integration constant. We assume a no-torque boundary
condition at the radius $x_{\rm ISCO}$ of the innermost stable circular
orbit, which is close to the BH; this requires that the
viscous angular-momentum current $\bfc_{\rm visc}=0$ at $x_{\rm ISCO}$
and from the first of equations (\ref{eq:cldef}) this in turn requires
$y=0$ at $x_{\rm ISCO}$. Thus
\begin{equation}
y(x)=\frac{c_M}{Q_1x^2}(x^{1/2}-x_{\rm ISCO}^{1/2}).
\label{eq:isco}
\end{equation}
We assume that the inner boundary of our integration region $x_a$ is
much larger than $x_{\rm ISCO}$ so in the region of interest
\begin{equation}
y(x)=\frac{c_M}{Q_1x^{3/2}},
\label{eq:ggg}
\end{equation}
which provides one boundary condition at $x=x_a$. 

At the outer radius $x_b$ the disc should lie in the plane of the
companion-star orbit, as we would expect if the disc is fed by mass
loss from the companion. Thus $\bfn=\bfn_\star$ at $x=x_b$, which
provides three additional boundary conditions. Moreover since
$|\bfn|=1$ at all radii, we must have $\bfn\cdot\p\bfn/\p x=0$ at
$x=x_b$, which provides another boundary condition (it is
straightforward to show from the second of eqs.\ \ref{eq:simple}
that these conditions are sufficient to ensure that $|\bfn|=1$ at all
radii). Note that we do not require that the disc lies in the equator
of the central body for $x\ll1$, although it turns out to do so in all of our
numerical solutions. 

Let us assume for simplicity that (i) inside the inner integration
boundary $x_a$ the external torques on the right side of the second of
equations (\ref{eq:simple}) vanish; (ii) the disc normal $\bfn$ is
nearly constant, $\bfn(x)=\bfn_0+\epsilon\bfn_1(x)$ where
$\epsilon\ll1$. Then to first order in $\epsilon$ the first of
equations (\ref{eq:simple}) is the same as for a flat disc, yielding
the solution (\ref{eq:ggg}). Substituting this result into the second
of equations (\ref{eq:simple}) and working to first order in
$\epsilon$ we find
\begin{equation}
\frac{d^2\bfn_1}{dx^2}
+\frac{3}{2x}\frac{d\bfn_1}{dx}=0\quad\mbox{with solution} \quad
\bfn_1=\bfa + \bfb x^{-1/2}
\label{eq:ibc}
\end{equation}
where $\bfa$ and $\bfb$ are constants.  To avoid an unphysical
solution that grows as $x\to 0$ we must have $\bfb=0$. The component
of $\bfb$ along $\bfn$ is already guaranteed to be zero because our
earlier boundary conditions ensure that $\bfn\cdot d\bfn/dx=0$. Thus
the two components of $d\bfn/dx$ perpendicular to $\bfn$ must vanish
at the inner boundary $x_a$, which provides the final two boundary
conditions. Note that there is no similar requirement at the outer boundary, since
the parasitic solution $\bfb x^{-1/2}$ decays as $x\to\infty$. 

The resulting boundary-value problem is solved using a collocation
method with an adaptive mesh (routine \textsc{d02tvf} from Numerical Algorithms
Group). To improve convergence we start with zero obliquity and
increase the obliquity in steps of $1^\circ$, using the converged
solution from each value of the obliquity as the initial guess for the
solution for the next. 

\begin{figure}
\centerline{\includegraphics[width=0.8\textwidth,bb=1 150 600 700]{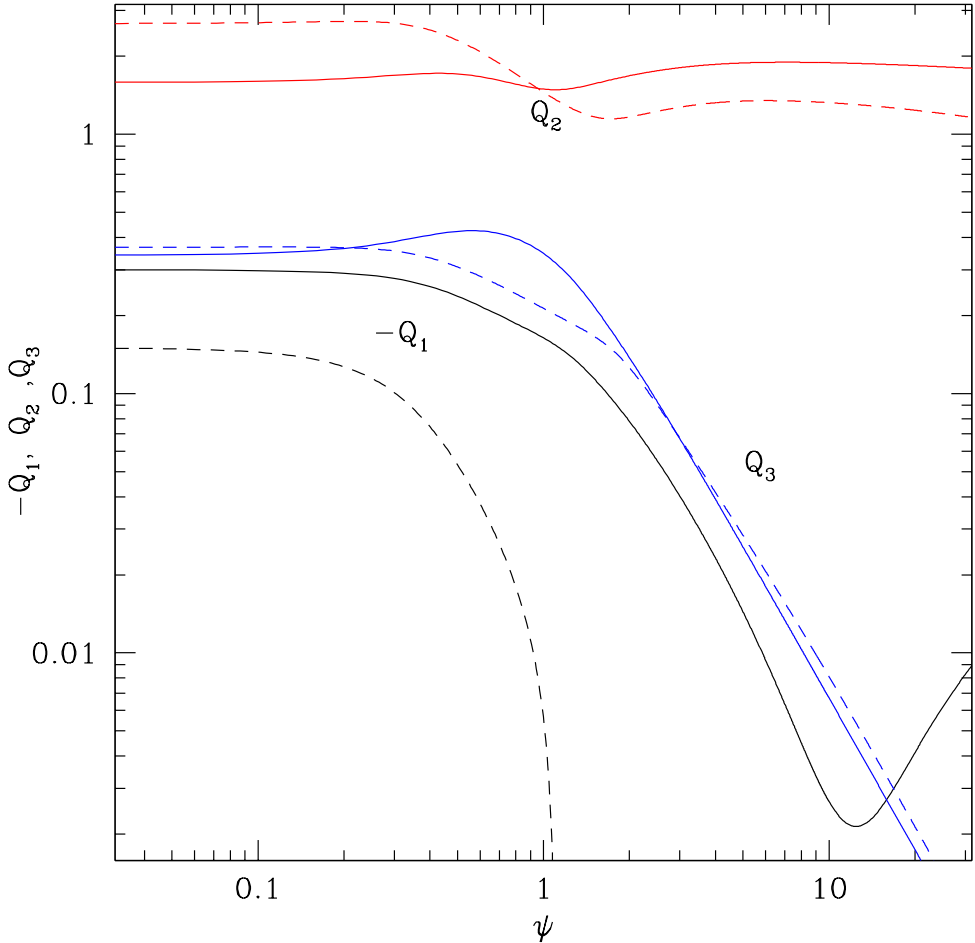}}
\caption{The viscosity coefficients $-Q_1$, $Q_2$, $Q_3$ for an
  isothermal disc with viscosity described by a Shakura--Sunyaev
  $\alpha$-model (eq.\ \ref{eq:alpha}) having $\alpha=0.2$,
  $\alpha_b=0$ (solid lines) or $\alpha=0.1$, $\alpha_b=0.1$ (dashed
  lines).  The horizontal coordinate is the dimensionless warp
  $\psi\equiv r|d\bfn/dr|$. We plot $-Q_1$ because $Q_1$ is normally
  negative for small warps; for $\alpha=0.2$, $\alpha_b=0$ $Q_1$ is
  negative for all $\psi$ while for $\alpha=0.1$, $\alpha_b=0.1$ $Q_1$
  is positive for $\psi>1.106$. The calculations follow the precepts
  of Ogilvie (1999) and employ a code provided by G.\ Ogilvie. }
\label{fig:two}
\end{figure}

\paragraph*{Time-dependent discs} We have solved the partial
differential equations (\ref{eq:ogthree}), typically over the interval
$[x_a,x_b]$ with $x_b=1/x_a=30$. Usually the viscosity coefficients
$Q_i$ are chosen to be appropriate for a disc with $\alpha=0.2$, $\alpha_b=0$. The
coefficients are determined as functions of the warp $\psi\equiv
x|\upartial\bfn/\upartial x|$ using a code generously provided by G.\
Ogilvie (see Fig.\  \ref{fig:two}); the coefficients are tabulated on a grid $0\le\psi\le10$ and
interpolated using cubic splines. Mass, and the corresponding angular
momentum for circular orbits, are added
at a constant rate with a Gaussian distribution in radius centred at
$x=10$ (i.e., well outside the warp) and the disc is followed until it
reaches a steady state. The integration is carried out using the
routine \textsc{d03pcf} from Numerical Algorithms Group. A complication is
that the dependence of the coefficient $Q_1$ on $\psi$ means that
equation (\ref{eq:ogthree}) is third-order in the spatial derivative;
to reduce this to a second-order equation we treat the mass current
$c_M$ as a fourth dependent variable in addition to the three
components of the angular momentum $\bfL$ and integrate the second of
equations (\ref{eq:cldef}) along with equations
(\ref{eq:ogthree}). 

As in the steady-state case we assume that the disc is aligned with
the companion-star orbit at large radii, so $\bfn=\bfn_\star$ at the
outer boundary $x=x_b$. We also assume that the steady-state relation
(\ref{eq:ggg}) between the surface density and the mass current in a
flat disc applies at the inner boundary $x_a$; this is plausible since
we expect the disc to achieve an approximate steady-state most rapidly
at small radii. We assume that there is an outer disc boundary $x_o\gg
x_b$ at which a no-torque boundary condition applies. In the
steady-state disc, arguments analogous to those leading to equations
(\ref{eq:qqwwrr})--(\ref{eq:ggg}) imply
\begin{equation}
y(x)=-\frac{c_M}{Q_1x^2}(x_o^{1/2}-x^{1/2}).
\end{equation}
This implies in turn that at the outer boundary
\begin{equation}
y(x_b)=-\frac{c_M}{Q_1x_b^2}(x_o^{1/2}-x_b^{1/2})\quad \mbox{and}\quad \bfc_L=c_Mx_o^{1/2}\bfn_\star .
\end{equation}
Typically we use $x_o=10x_b$. 
Finally, the angular-momentum current at $x_{\rm ISCO}$ is $\bfc_L=x^{1/2}_{\rm
  ISCO}c_M\bfn$ which can be taken to be zero since $x_{\rm ISCO}$ is
very small. Since the disc is flat
inside the warp radius and the inner integration boundary $x_a$ is
much less than the warp radius, we may assume that $\bfc_L$ is
constant between $x_{\rm ISCO}$ and $x_a$ so we set $\bfc_L(x_a)=0$.

We usually start with a low-density disc and zero obliquity, and add
mass and angular momentum outside the warp radius at a constant 
rate until the disc reaches a steady state; then we slowly increase
the obliquity to the desired value. 

\subsection{Results}

\label{sec:results}

\begin{figure}
\includegraphics[width=0.5\textwidth]{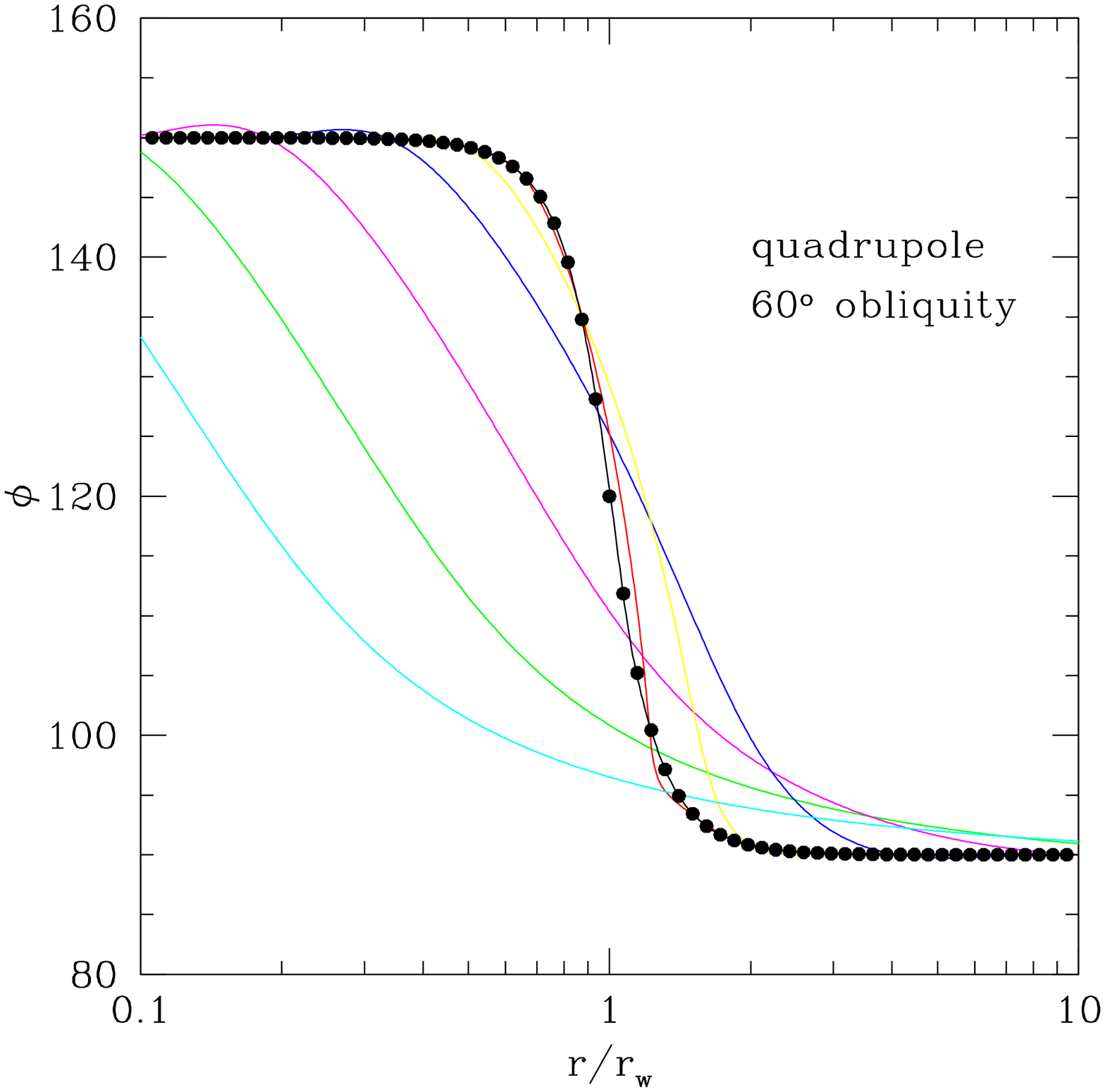}
\includegraphics[width=0.5\textwidth]{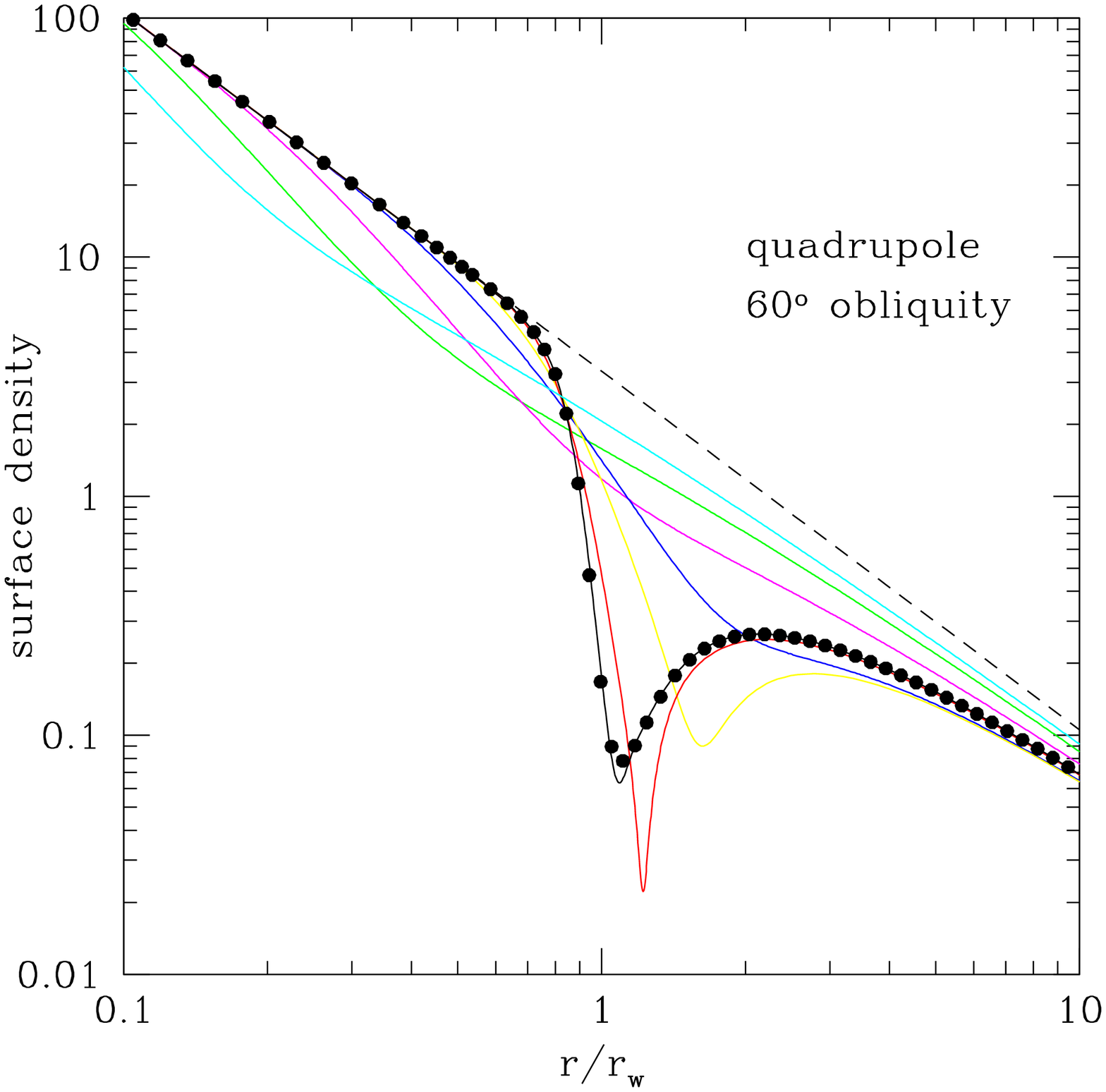}
\caption{(left) The orientation of a stationary disc orbiting a planet
  that has an obliquity of $60^\circ$ (from eqs. \ref{eq:simple}). The
  viscosity coefficients are $Q_1=-0.3$, $Q_2=1.58416$, appropriate
  for a flat disc with $\alpha=0.2$, $\alpha_b=0$, and the mass current is
  $c_M=-1$. The solutions shown have the parameter $\beta$ (eq.\
  \ref{eq:qdef}) representing the ratio of viscous torques to external
  torques equal to 1000 (cyan), 100 (green), 10 (magenta), 1 (blue),
  0.1 (yellow), 0.01 (red), 0.001 (black). The solid black circles
  represent the inviscid solution (the Laplace surface), given by
  equation (\ref{eq:lap}) and shown in the left panel of Fig.\ 
  \ref{fig:one}. (right) The surface density $y(x)$ for the discs
  shown in the left panel. The solid circles show the solution given
  by the first of equations (\ref{eq:simple}) and the orientation
  $\bfn(x)$ of the inviscid disc. The dashed line shows the surface
  density for a flat disc (eq.\ \ref{eq:ggg}).}
\label{fig:three}
\end{figure}

\paragraph*{Quadrupole discs}
The left panel of Fig.\  \ref{fig:three} shows the solutions of equation
(\ref{eq:simple}) for a planet obliquity of $60^\circ$ and a range of
viscosity parameters $\beta$ from 1000 to 0.001. As one might expect, very viscous discs
($\beta\gg1$) exhibit a smooth, gradual warp while low-viscosity discs
($\beta\ll1$) are close to the inviscid disc (eq.\ \ref{eq:lap}),
shown as the solid circles. 

The right panel shows the surface density $y(x)$. Here the behavior is
more interesting. While the surface density in very viscous discs is
close to that of a flat disc (dashed line, from eq.\ \ref{eq:ggg}), as
the viscosity is lowered the disc develops a sharp valley -- almost two
orders of magnitude -- in the surface density near the warp radius
$r_w$. The valley presumably occurs because the viscous stresses are larger when
the warp $\psi=x|d\bfn/d x|$ is large, so the mass and
angular-momentum current can be carried by a smaller surface
density. The asymptotic behavior of the surface density as the
viscosity becomes small is obtained from the first of equations
(\ref{eq:simple}) by substituting for $|d\bfn/dx|$ the value from the
inviscid solution (\ref{eq:lap}); this is shown as the solid circles
in the right panel of Fig.\  \ref{fig:three}.

\begin{figure}
\centerline{\includegraphics[width=0.8\textwidth,bb= 0 140 575 700]{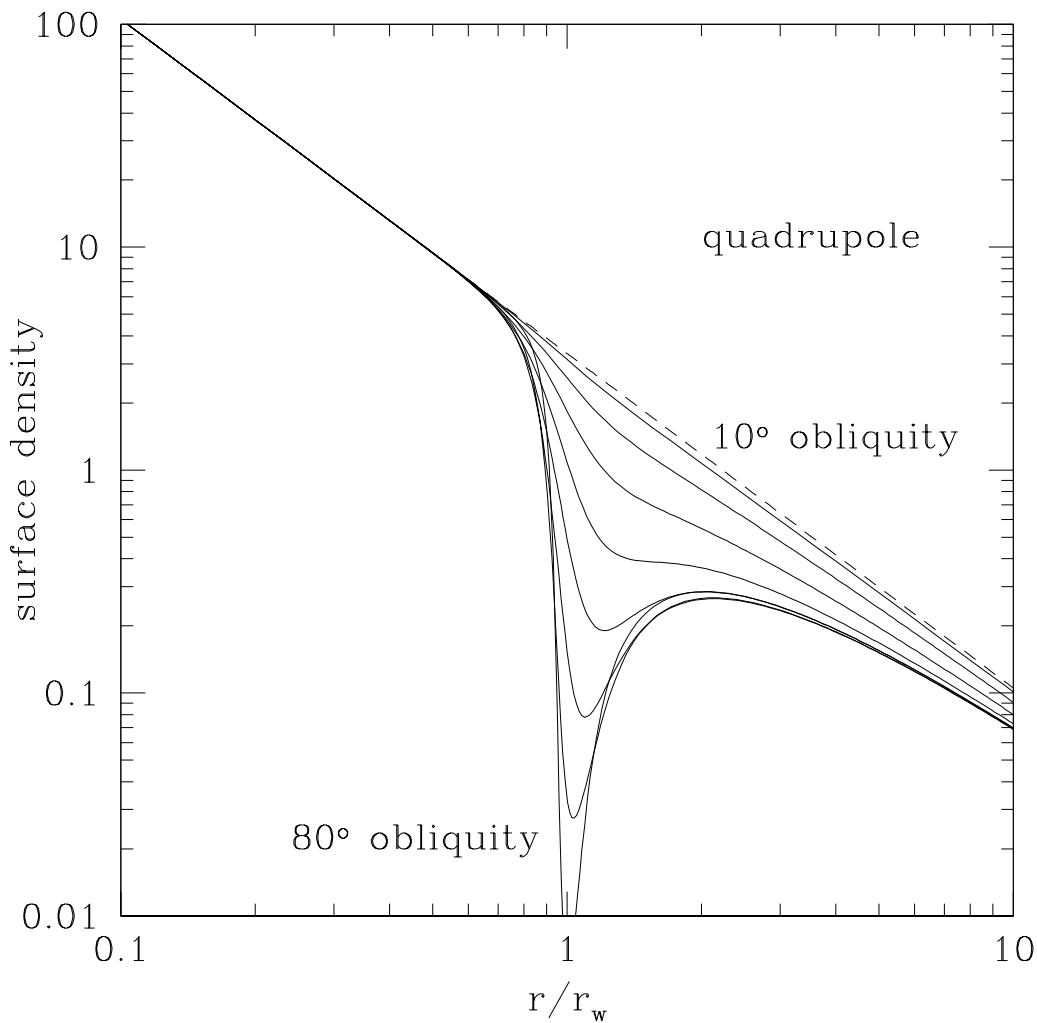}}
\caption{The surface density $y(x)$ for quadrupole discs with
  negligible viscosity ($\beta\to0$) and obliquity
  $10^\circ, 20^\circ,\ldots,80^\circ$. The other parameters of the discs
  are the same as in Fig.\  \ref{fig:three}. The dashed line shows the
  surface density for a flat disc (eq.\ \ref{eq:ggg}).}
\label{fig:four}
\end{figure}

The nature of the surface-density valley associated with the warp is
illustrated further in Fig.\  \ref{fig:four}, which shows the
surface-density profile for low-viscosity discs ($\beta\to0$) for
obliquities $10^\circ,20^\circ,\ldots,80^\circ$. As the obliquity
grows the valley becomes deeper: at an obliquity of $80^\circ$ the
surface density is only 0.2 per cent of the surface density in an unwarped
disc at the bottom of the valley, near radius $1.00r_w$. 

\begin{figure}
\includegraphics[width=0.5\textwidth]{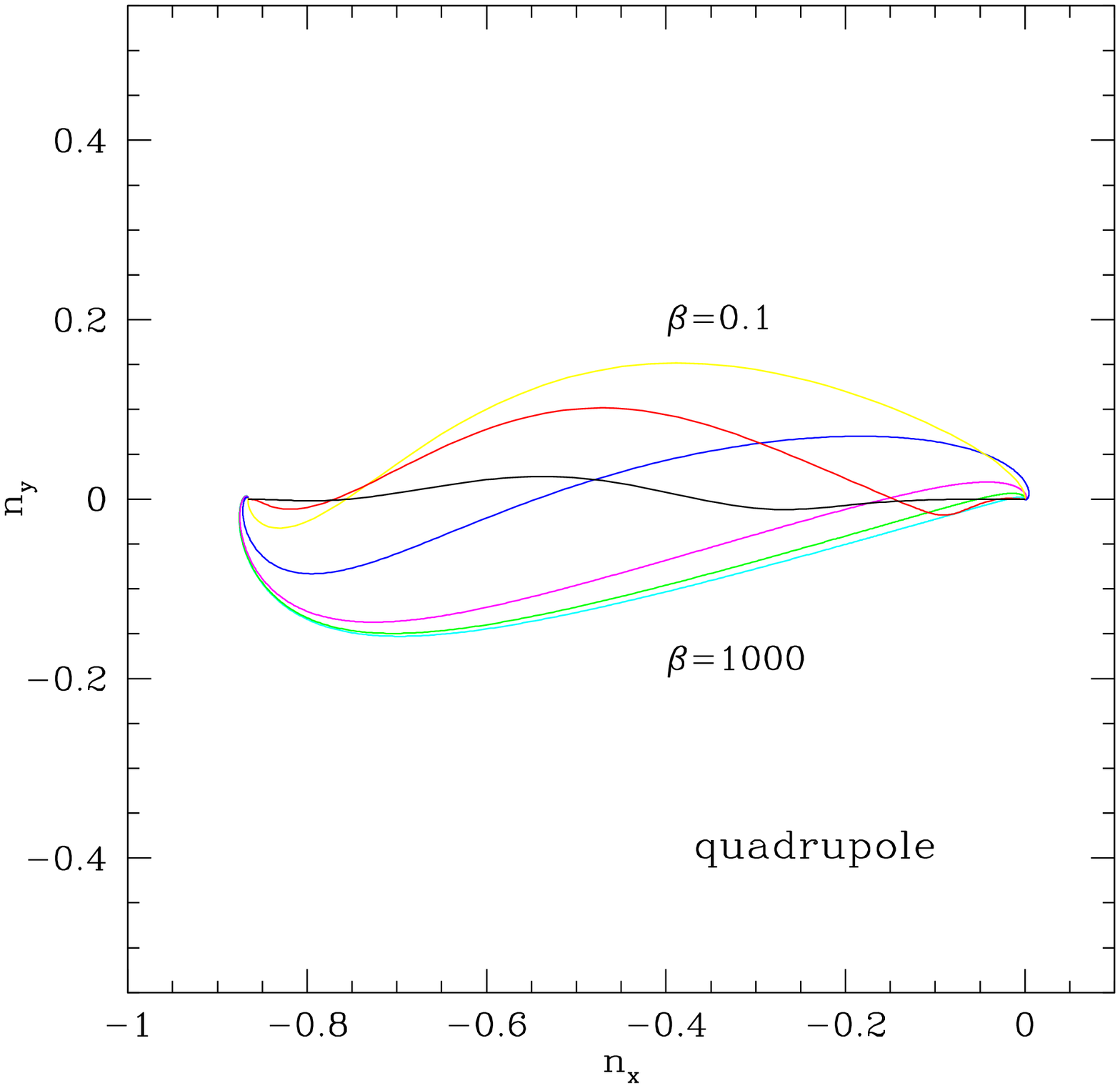}
\includegraphics[width=0.5\textwidth]{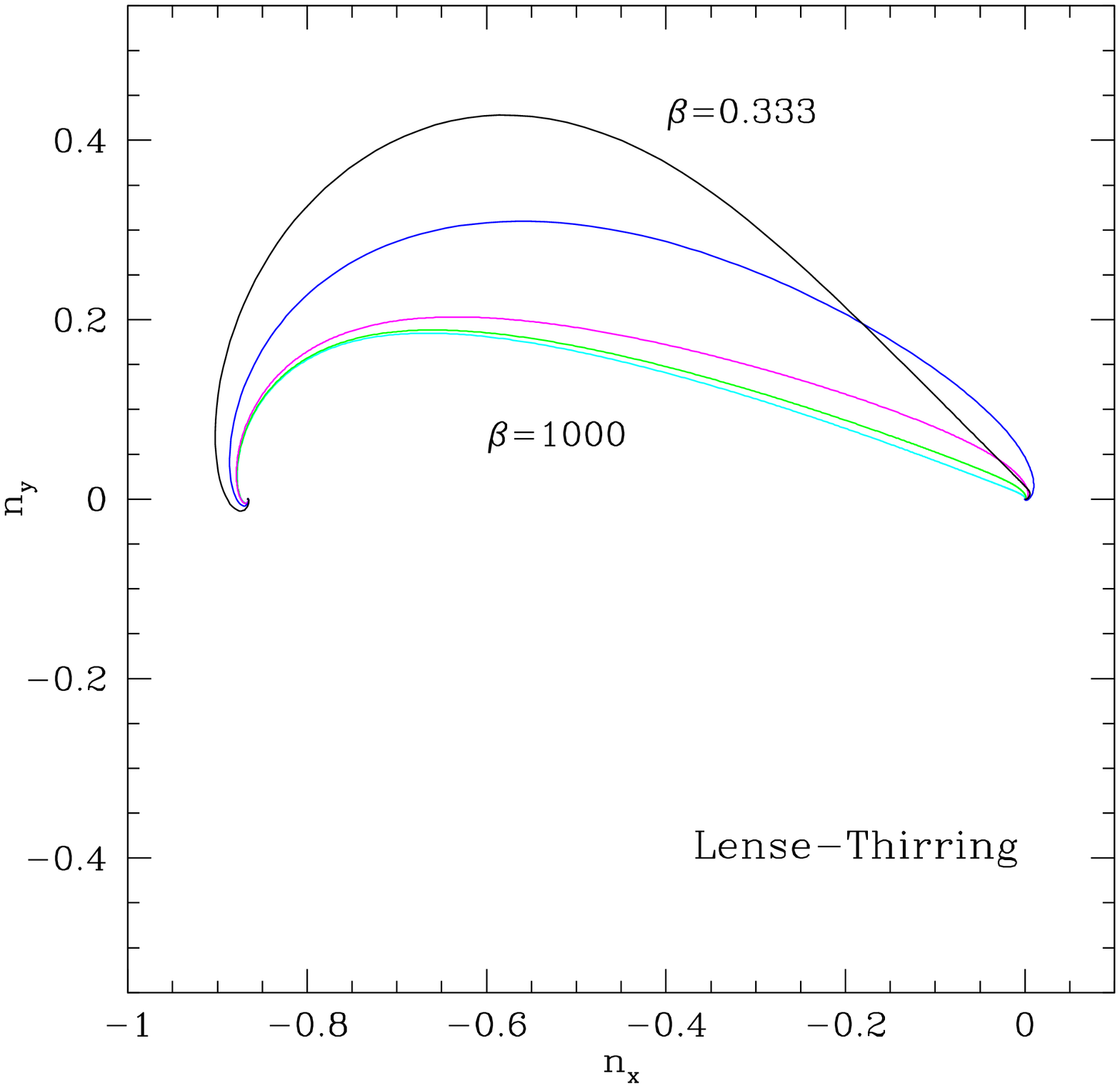}
\caption{The horizontal components $(n_x,n_y)$ of the unit normal
  vector for quadrupole discs (left panel) and Lense--Thirring discs
  (right panel). The obliquity is $60^\circ$ and the other parameters
  are as described in Fig.\  \ref{fig:three} (left panel) or
  \ref{fig:five} (right panel). In both panels the parameter $\beta$ (eq.\
  \ref{eq:qdef}), representing the ratio of viscous torques to external
  torques, is equal to 1000 (cyan), 100 (green), 10 (magenta), 1 (blue);
  in the left panel there are additional curves for $\beta= 0.1$
  (yellow), 0.01 (red), 0.001 (black) and in the right panel there is
  an additional curve for the critical value $\beta=0.333$ (black).}
\label{fig:foura}
\end{figure}

The steady-state warped discs also exhibit some spirality or twisting;
this is shown in Fig.\  \ref{fig:foura} by plotting the horizontal
components $(n_x,n_y)$ of the unit vector normal to the disc. 

\begin{figure}
\includegraphics[width=0.5\textwidth]{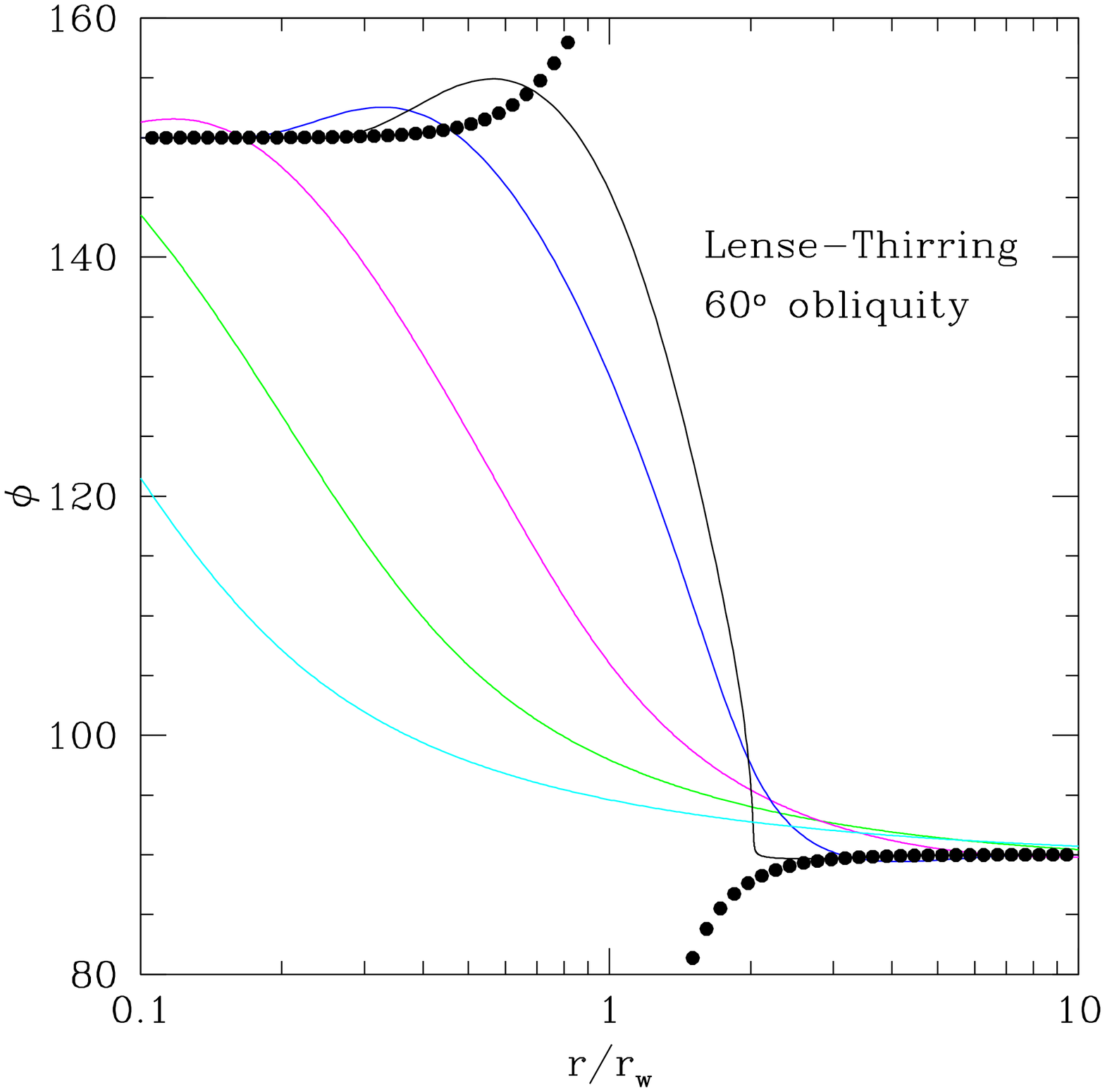}
\includegraphics[width=0.5\textwidth]{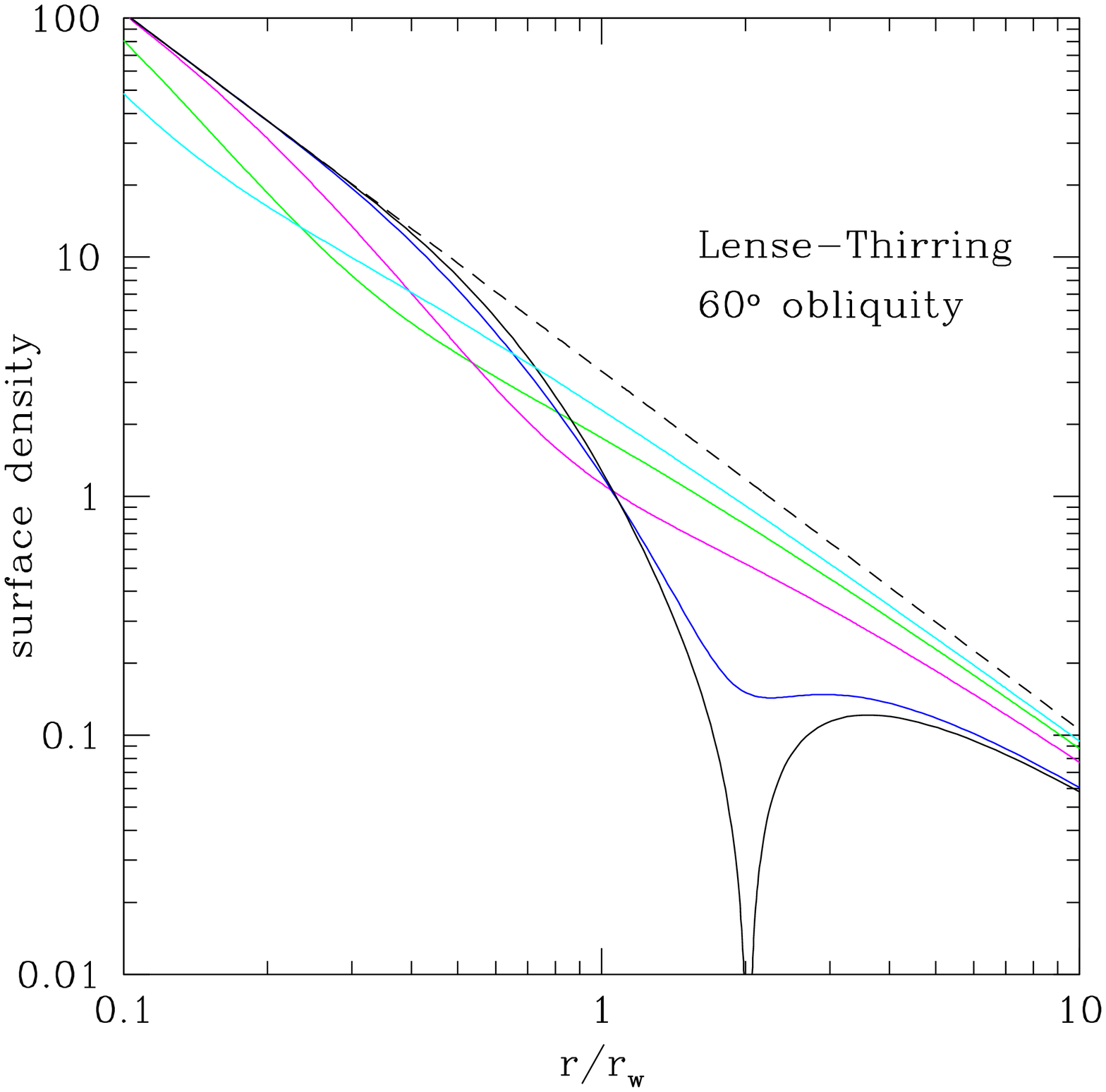}
\caption{(left) The orientation of a stationary disc orbiting a BH
  that has an obliquity of $60^\circ$ (from eqs. \ref{eq:simple}). The
  parameters are the same as in Fig.\  \ref{fig:three}, except that
  the parameter $\beta$ (eq.\ \ref{eq:qdef}) representing the ratio of
  viscous torques to external torques equals 1000 (cyan), 100 (green),
  10 (magenta), 1 (blue), and 0.333 (black). For $\beta<0.333$ no
  solution exists. The solid black circles represent the inviscid
  solution, given by equation (\ref{eq:kerr}) and shown in the right
  panel of Fig.\  \ref{fig:one}. (right) The surface density $y(x)$
  for the discs shown in the left panel. The dashed line shows the
  surface density for a flat disc (eq.\ \ref{eq:ggg}).}
\label{fig:five}
\end{figure}

\paragraph*{Lense--Thirring discs}
Fig.\  \ref{fig:five} is analogous to Fig.\  \ref{fig:three}: it shows
the solutions of equation (\ref{eq:simple}) for a Lense--Thirring disc when the
BH obliquity is $60^\circ$. The viscosity parameter $\beta$
ranges from 1000 to 0.333; for $\beta<0.333$ no steady-state solution
exists. Similarly, the right panel of Fig.\  \ref{fig:five} shows the
horizontal components of the unit normal in Lense--Thirring discs with
$60^\circ$, to be compared with the left panel of the same figure for
quadrupole discs. 

The absence of steady-state solutions for Lense--Thirring discs for viscosity
less than some critical value at fixed obliquity -- or obliquity larger
than a critical value at fixed viscosity -- is a novel feature not seen in the
quadrupole discs, and presumably related to the jump seen in the
orientation of inviscid Lense--Thirring discs (\S\ref{sec:invisc}). 

\begin{figure}
\centerline{\includegraphics[width=0.8\textwidth,bb=0 150 580 700]{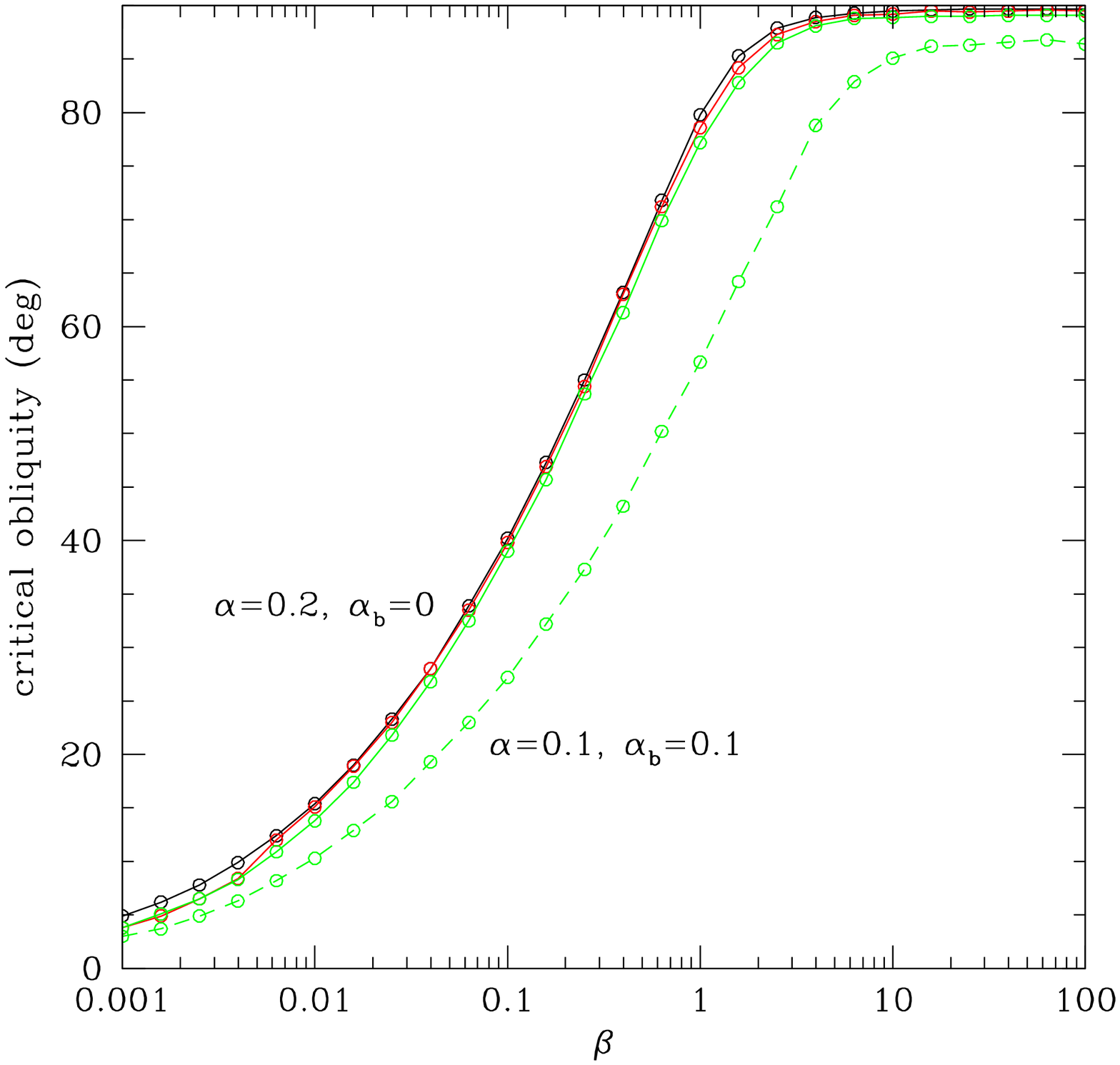}}
\caption{Above the critical obliquity shown here, steady-state
  Lense--Thirring disc solutions do not exist. The parameter $\beta$
  measures the strength of the viscous forces (eq.\
  \ref{eq:qdef}). The solid lines are for Shakura--Sunyaev discs with
  $\alpha=0.2$, $\alpha_b=0$ and the dashed line is for
  $\alpha=\alpha_b=0.1$. The black and red curves are derived from
  steady-state and time-dependent disc models (eqs.\ \ref{eq:ogthree}
  and \ref{eq:simple}) with the viscosity parameters $Q_1$ and $Q_2$
  set to their unwarped values and $Q_3=0$. The green curves are for
  $Q_i$ depending on the local warp, as in Fig.\  \ref{fig:two}.
}
\label{fig:six}
\end{figure}

Fig.\  \ref{fig:six} illustrates how the critical obliquity and
viscosity parameter are related. The black  curve shows the
critical values for the simplified steady-state equations
(\ref{eq:simple}), with $Q_1=-0.3$, $Q_2 = 1.58416$, $Q_3=0$. The
critical values are defined here by the point where the maximum warp
$\psi=10$; this is generally close to the curve with $\psi\to\infty$
and for $\psi\ga 10$ it is unlikely that our model is accurate in
any case.

The red curve in Fig.\  \ref{fig:six} shows the critical values
obtained by solving the time-dependent equations (\ref{eq:ogthree})
for the same constant values of $Q_i$; in this case the critical
values are defined by the obliquity at which the maximum warp of the
time-dependent solution exceeds $\psi=10$.  The agreement of the red
and black curves is partly a successful check of our steady-state
and time-dependent numerical codes, but more importantly it implies
that time-dependent discs with obliquity above the critical value will
develop singular warps -- that is, for example, there is no oscillating
solution of the time-dependent Pringle--Ogilvie equations that remains non-singular. 

The green curve shows the critical values obtained from equations
(\ref{eq:ogthree}) with viscosity parameters $Q_i$ that depend on the
warp as shown in Fig.\  \ref{fig:two}. This exhibits the same
qualitative behavior as the black and red curves, demonstrating that
the critical values are not strongly dependent on the variation of
viscosity parameters with the strength of the warp.

Finally, the green dashed curve is the same as the green solid curve,
but for parameters $Q_i$ appropriate for Shakura--Sunyaev parameters
$\alpha=0.1$, $\alpha_b=0.1$.

What happens to a Lense--Thirring accretion disc when the obliquity
exceeds the critical value is not understood. Finite-time
singularities (`blow-up') are a common feature of non-linear parabolic
partial differential equations such as the Pringle--Ogilvie equations
and it is likely that the absence of a solution reflects the
approximation of the correct, hyperbolic, fluid equations with
diffusion equations. The limitations of the diffusion approximation in
warped discs are well-known: \cite{pp83} argue that a transition from
diffusive to wavelike behavior occurs when $\alpha$ decreased below
$H/r$ (see also \citealt{pl95} and
\citealt{og06}). In this regime, bending waves governed by the
pressure in the disk could transport angular momentum to connect
smoothly the inner and outer disks. The behavior of such waves in
Lense--Thirring discs is described by \cite{lop02} but only to linear
order in the warp amplitude, where the singular behavior is not
present. For finite-amplitude warps, it is far from clear how
to incorporate the required extra physics into the Pringle--Ogilvie
equations or what behavior we might expect.

The sharp changes in disc orientation seen in Fig.\  \ref{fig:five}
are reminiscent of the phenomenon of `breaking' in which the
orientation of the accretion disc changes almost discontinuously
\citep{nk12,nix12}, although there are substantial differences in the
phenomenology and interpretation (see \S\ref{sec:other} for further
discussion). 

\subsection{The behavior of the disc at the critical obliquity}

\label{sec:critical}

\noindent
At the critical obliquity or viscosity there is a radius (the
`critical radius') at which the surface density approaches zero and
the disc warp $\psi=r|d\psi/dr|$ changes from near zero to a very
large value (black curves in Fig.\  \ref{fig:five}). We can offer some analytic
insight into this behavior.

Since the behavior of the disc changes sharply in a small radial
distance, this change is unlikely to be due to the external torques, which
vary smoothly with radius. Thus we examine the governing differential
equations (\ref{eq:ogthree}) with the right-hand side and
$\p/\p\tau$ set to zero. Then this equation states that the total angular-momentum current 
$\bfc_{\rm visc}+x^{1/2}c_M\bfn$ must be independent of radius $x$. We
erect a coordinate system specified by the triple of unit vectors
$\bfe_1,\bfe_2,\bfe_3$ with $\bfe_3$ parallel to the angular-momentum
current, so $\bfc_{\rm visc}+x^{1/2}c_M\bfn=c_L\bfe_3$ with the mass
and angular-momentum currents $c_M$ and $c_L$ constants. For
simplicity we assume that the viscosity coefficients $Q_1$,
$Q_2$ are constants, and $Q_3=0$. Then
\begin{equation}
x^{1/2}c_M\bfn -Q_1x^2y(x)\bfn
-Q_2x^3y(x)\frac{d\bfn}{dx}=c_L\bfe_3.
\label{eq:one}
\end{equation}
Since $\bfn$ is a unit vector, $\bfn\cdot d\bfn/dx=0$,
we may take the dot product with $\bfn$ to obtain
\begin{equation}
x^{1/2}c_M -Q_1x^2y(x)=c_Lf(x) \quad\mbox{where}\quad
f(x)\equiv\bfn\cdot\bfe_3=n_3.
\label{eq:two}
\end{equation}
The components of (\ref{eq:one}) along $\bfe_1$ and $\bfe_2$ are
\begin{equation}
\left[x^{1/2}c_M -Q_1x^2y(x)\right]n_{1,2} - Q_2x^3y(x)\frac{dn_{1,2}}{dx}=0.
\label{eq:three}
\end{equation}
Combining equations (\ref{eq:two}) and (\ref{eq:three}) with the
conditions $\sum_{i=1}^3 n_i^2=1$, $\sum_{i=1}^3 n_i dn_i/dx=0$, we
find
\begin{equation}
\frac{df}{dx}=\frac{Q_1}{Q_2x}\frac{1-f^2}{f-x^{1/2}c_M/c_L}.
\label{eq:ode}
\end{equation}
The interesting behavior occurs if the mass and angular-momentum
current have the same sign. In this case the non-linear differential
equation (\ref{eq:ode}) has a critical point at $f=1$,
$x=(c_L/c_M)^2\equiv x_c$.  If we restrict ourselves to the usual case
in which $Q_1<0$, $Q_2>0$, then near the critical point solutions must
take one of the following two forms:

\begin{enumerate}

\item $f=1$; this implies an unwarped disc with normal parallel to the
  angular-momentum current. The surface density is given by equation
  (\ref{eq:two}) as
\begin{equation}
y(x)=\frac{c_M}{2Q_1x_c^{5/2}}(x-x_c) + \mbox{O}(x-x_c)^2.
\label{eq:kkpp}
\end{equation}
In the usual case where the mass current $c_M<0$ this
solution is physical (positive surface density) for $x>x_c$, i.e.,
outside the critical point. 

\item In this case
\begin{equation}
f(x)=1+\frac{Q_2-4Q_1}{2Q_2x_c}(x-x_c) + \mbox{O}(x-x_c)^2, \qquad
y(x)=\frac{2c_M}{Q_2x_c^{5/2}}(x-x_c) + \mbox{O}(x-x_c)^2.
\label{eq:hhyy}
\end{equation}
Since $f<1$ and $y>0$ this solution is only physical when the mass
current $c_M<0$ and then only for $x<x_c$, i.e., inside the critical
point. The angle between the angular momentum current and the disc
normal is $\theta$ where $\cos\theta=f$ so $\theta\sim (x_c-x)^{1/2}$ and the warp
$\psi=x|d\bfn/dx|\sim (x_c-x)^{-1/2}$. Thus the warp angle $\psi$ is singular at
the critical point. 

\end{enumerate}

The behavior of these solutions is consistent with the behavior seen
in Fig.\  \ref{fig:five} at the critical obliquity: outside the
critical radius, the disc is flat and the surface density decreases
linearly to zero as the radius decreases to the critical radius (eq.\ \ref{eq:kkpp}), while
inside the critical radius the azimuthal angle $\phi-\half\upi$ of the warp
normal varies as $(x_c-x)^{1/2}$, and the surface density decreases
linearly to zero as the radius increases to the critical radius (eq.\
\ref{eq:hhyy}). Since the surface density is zero at the critical
point, there is no viscous angular-momentum transport across it, only
advective transport.

\section{Evolution of viscous discs with self-gravity}

\label{sec:sg}

\noindent
Our treatment of accretion discs with self-gravity will be briefer and
more approximate than the treatment of discs with a companion in the
preceding section, for three main reasons: (i) AGN accretion discs are
the only ones in which self-gravity is likely to be important, and
these are less well-understood than accretion discs around
stellar-mass BHs; (ii) the theory of bending waves in 
gas discs is remarkably sensitive to small deviations from Keplerian
motion (cf.\ eq.\ \ref{eq:nonres}); (iii) we found that warped steady-state accretion
discs around a spinning BH with a companion do not exist for some
values of the obliquity and viscosity, and this finding requires the
best available disc models to be credible. In contrast we shall find
that warped discs with self-gravity exhibit interesting but physically
plausible behavior even in relatively simple disc models, and there is
no reason to believe that this behavior will change qualitatively in
more sophisticated treatments.

We shall assume that the warp is small so that linearized theory can
be used, and that the disc surface-density distribution is the same as
in a flat disc. We shall also assume a simple model for the viscous
damping of the warp. 

We also ignore the effects of pressure in the disc. This assumption is
problematic because \cite{pl95} showed that in gravitationally stable
Keplerian discs ($Q>1$ in eq.\ \ref{eq:toomredef}) the dispersion
relation for bending waves is dominated by pressure rather than
self-gravity. However, (i) this result depends sensitively on whether the
disc is precisely Keplerian, and small additional effects such as
centrifugal pressure support or relativistic apsidal precession can
dramatically reduce the influence of pressure on the dispersion
relation; (ii) modifying the Pringle--Ogilvie equations to include
pressure is a difficult and unsolved problem. 

The normal to the disc at radius $r$ is $\bfn=(n_x,n_y,n_z)$. We
choose the axes so that the BH spin is along the positive $z$-axis;
then since the warp is small $|n_x|,|n_y|\ll 1$. Write
$\zeta(r,t)\equiv n_x+{\rm i}n_y$; then neglecting all terms quadratic in
$\zeta$ the Lense--Thirring torque (\ref{eq:lt}) causes precession of
the angular momentum at a rate
\begin{equation}
\frac{d\zeta}{dt}(r,t)\bigg|_{\rm
  LT}=\frac{2(GM_\bullet)^2a_\bullet}{c^3r^3}\,{\rm i}\zeta(r,t).
\end{equation}

The equations of motion due to the self-gravity of the warped disc are
given by classical Laplace--Lagrange theory \citep{md99},
\begin{equation}
\frac{d\zeta}{dt}(r,t)\bigg|_{sg}=-\frac{{\rm i}\upi G}{2(GM_\bullet
  r)^{1/2}}\int \frac{r'\,\Sigma(r')\,dr'}{\mbox{max\,}(r,r')}\,\chi
b_{3/2}^{(1)}(\chi)[\zeta(r,t)-\zeta(r',t)]
\end{equation}
where $\Sigma(r)$ is the surface density,
$\chi=\mbox{min}\,(r,r')/\mbox{max}\,(r,r')$ and the Laplace
coefficient 
\begin{equation}
b_{3/2}^{(1)}(\chi)=\frac{2}{\upi}\int_0^\upi\frac{\cos
  x\,dx}{(1-2\chi\cos x
  +\chi^2)^{3/2}}=\frac{4}{\upi\chi(1-\chi^2)^2}[(1+\chi^2)E(\chi)-(1-\chi^2)K(\chi)]
\end{equation}
with $K(\chi)$ and $E(\chi)$ complete elliptic integrals.

The equations of motion due to viscosity are derived by simplifying
equations (\ref{eq:ogtwo}) and (\ref{eq:ogfour}). The angular-momentum
current proportional to $Q_1\bfn$ and the mass current $C_M$ determine
the steady-state surface density in a flat disc, which we assume to be
given, so we drop these terms. The current proportional to $Q_3$
appears to play no essential role, so we drop this term as
well. Furthermore we assume that the sound speed $c_s$ is independent
of radius (isothermal disc), and we replace $Q_2$ by
$\frac{1}{2}\alpha_\perp$ (eq.\ \ref{eq:qalpha}). Thus we find
\begin{equation}
\frac{d\zeta}{dt}(r,t)\bigg|_v=\frac{c_s^2\alpha_\perp}{2(GM_\bullet
  r^3)^{1/2}\Sigma(r,t)}\frac{\p}{\p r}r^3\Sigma(r,t)\frac{\p \zeta}{\p r}.
\end{equation}
We now look for a steady-state solution in which
$d\zeta/dt|_{LT+sg+v}=0$. We replace the radius by the
dimensionless variable $x=r/r_w$ where $r_w$ is defined for a
self-gravitating disc by equation (\ref{eq:rwself}), and we assume that the surface density is a power
law, $\Sigma(r)=\Sigma_0 /x^s$. The equations above simplify to
\begin{equation}
\frac{4}{x^{5/2}}\zeta- \int \frac{{x'}^{1-s}\,dx'}{\mbox{max\,}(x,x')}\chi
b_{3/2}^{(1)}(\chi)[\zeta(x)-\zeta(x')] - {\rm i}\gamma \alpha_\perp x^{s-1}\frac{d}{dx}x^{3-s}\frac{d\zeta}{dx}=0
\label{eq:xxcc}
\end{equation}
where $\gamma$ is the viscosity parameter defined in equation
(\ref{eq:betaself}). We impose the boundary conditions $d\zeta/dx=0$ as
$x\to0$ and $x\to\infty$ (the disc is flat near the BH, and flat far outside
the warp radius) and $\zeta\to \zeta_0$ at $x\to\infty$ (at large distances
the normal to the disc is inclined to the spin axis of the BH by an
angle $\theta=|\zeta_0| \ll1$). Since equation (\ref{eq:xxcc}) is linear, there
is no loss of generality if we set $\zeta_0=1$.

In these dimensionless units, the shape of the warp is determined by
only two parameters, the logarithmic slope of the surface-density
distribution $s$, and the viscosity parameter $\gamma
\alpha_\perp$. The relation between $\alpha$ and $\alpha_\perp$ is
discussed after equation (\ref{eq:qalpha}).

\begin{figure}
\includegraphics[width=0.95\textwidth,bb=0 150 580 700]{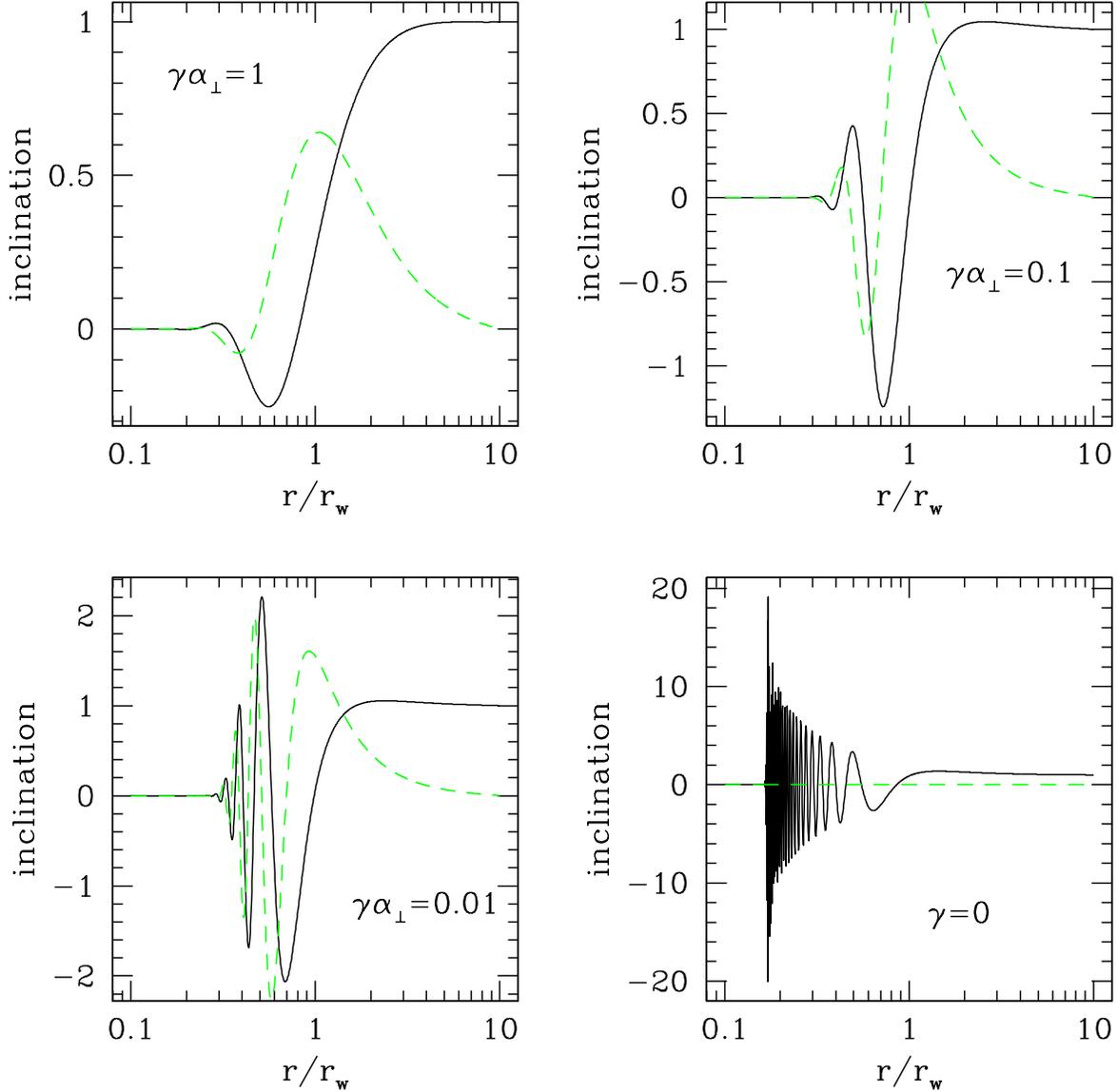}
\caption{The steady-state shape of warped discs including
  Lense--Thirring torque, self-gravity, and viscosity (eq.\
  \ref{eq:xxcc}). The four panels show four different values of the
  viscosity parameter $\gamma \alpha_\perp$ (eq.\
  \ref{eq:betaself}). The figures plot the real and imaginary parts of
  the complex inclination $\zeta$ (solid black and dashed green lines)
  as a function of the radius in units of the warp radius $r_w$ (eq.\
  \ref{eq:rwself}).  At large radii the disc is assumed to be flat
  with $\zeta=1$; since eq.\ (\ref{eq:xxcc}) is linear the results can
  be scaled to any (small) inclination. At small radii the disc is
  found to lie in the BH equator, $\zeta=0$. Note the different
  vertical scales in the four panels. The disappearance of the
  oscillations at $x<0.18$ in the lower right panel is a numerical
  artifact due to limited resolution.}
\label{fig:sg}
\end{figure} 

Fig.\  \ref{fig:sg} shows the solutions of equation (\ref{eq:xxcc})
for the surface-density slope $s=\frac{3}{5}$ appropriate for a
gas-pressure dominated disc (eq.\ \ref{eq:siggas}). The solid and
dashed lines show the real and imaginary parts of $\zeta(x)$. For
low-viscosity discs ($\gamma \alpha_\perp\ll1$) we find that the disc
develops bending waves inside the warp radius, and if the viscosity
is sufficiently small the bending waves can grow in amplitude by
orders of magnitude as the radius shrinks (the disappearance of the
bending waves at $x<0.18$ in the lower right panel is a numerical
artifact, which arises because the wavelength of the bending waves
becomes shorter than the resolution of the numerical grid,
$\Delta\log_{10}x=0.002$).

Many of the properties of the bending waves can be understood using a WKB
analysis \citep[][hereafter SCL83]{shu83}.  We shall quote the results from this paper
without derivations. If we assume that the waves have the form
$\zeta=A_\zeta(r)\exp[{\rm i}\Phi(r)]$ with radial wavenumber $k\equiv d\Phi/dr$, then
the dispersion relation is (SCL83 eq.\ 22, with $\omega=0$ and $m=1$)
\begin{equation}
|k|=\frac{2G^{3/2}M_\bullet^{5/2}a_\bullet}{\upi c^3\Sigma(r) r^{9/2}}.
\label{eq:wkb}
\end{equation}
The WKB approximation is valid if the waves have short wavelengths,
$|k|r\la 1$, which in turn requires that the radius is less than
the warp radius $r_w$ defined in equation (\ref{eq:rwself}); and this
in turn requires that the dimensionless variable $x$ in Fig.\ 
\ref{fig:sg} is small compared to unity. For plausible variations of
the surface density $\Sigma(r)$, the wavelength $2\upi/|k|$ gets
shorter and shorter as the radius shrinks.

In the absence of viscosity, the maximum inclination of the bending wave varies as
$A_\zeta(r)\propto [r^{3/2}\Sigma(r)]^{-1}$ (SCL83, eq.\ 34, with the
inclination amplitude $A_\zeta=A(r)/r$) so if the surface density falls as
$r^{-s}$ then the amplitude of the warp grows as the radius shrinks
whenever $s<\frac{3}{2}$, which is true for most disc models. 

The waves are spiral, as may be deduced from the offset between the
solid (real) and dashed (imaginary) curves in Fig.\  \ref{fig:sg}
(except in the lower right panel, where the viscosity is zero). The
dispersion relation (\ref{eq:wkb}) does not distinguish leading and
trailing waves but causality arguments do: trailing waves propagate
inward (i.e. negative group velocity, see SCL83 eq.\ 23) while leading
waves propagate outward. Waves excited by the warp in the outer part
of the disc and damped at small radii by viscosity must propagate
inward and hence are trailing.

In the case of low-viscosity Lense--Thirring discs that are warped
because of a companion, we found that no solutions of the
Pringle--Ogilvie equations existed above a critical obliquity. These
calculations suggest that self-gravitating discs are more
well-behaved -- that the long-range nature of the gravitational force
allows a smooth transition from the outer to the inner orientation for
any viscosity and obliquity, through the excitation of bending waves
that are eventually damped by viscosity as they propagate
inward. However, we caution that the analysis of this section is
linear in the warp amplitude and it is possible that non-linear effects
will prohibit a continuously varying warp shape once the obliquity is
large enough. 

This physical picture needs to be modified for AGN discs dominated by
radiation pressure, where the surface density varies as
$\Sigma(r)\propto r^{3/2}$ (eq.\ \ref{eq:sigrad}) out to a radius
$r_{pr}$ (eq.\ \ref{eq:rpr}) where gas pressure begins to dominate,
after which the surface density declines as $r^{-3/5}$. If $r_{pr}\la
r_w$, the bending waves are launched as usual at the warp radius $r_w$
and propagate smoothly into the region $r<r_{pr}$, although their
dispersion relation will change once they enter the
radiation-dominated region. If $r_{pr}$ is larger than $r_w$, the
gravitational torque will include a significant contribution from
material in the accretion disc near $r_{pr}$ (the torque from material
between $R\gg r$ and $2R$ varies as $G\Sigma(R)r^2/R\sim R^{1/2}$) in
addition to the gravitational torque from local material. This extra
torque will tend to counter-act the Lense--Thirring torque, and if it
is large enough will prevent the excitation of bending waves.

In summary, for low-viscosity discs in which self-gravity is
important, misalignment of the disc axis at large radii with the BH spin
axis can excite bending waves inside the warp radius
(\ref{eq:rwself}). For discs dominated by gas pressure, where the
surface density $\Sigma(r)\propto r^{-0.6}$, Fig.\  \ref{fig:sg} shows
that the condition for exciting oscillatory waves is $\gamma
\alpha_\perp\simeq 0.05$. For warps of sufficiently small amplitude,
$\alpha_\perp=\frac{1}{2}\alpha^{-1}$ (eq.\ \ref{eq:qalpha}) so the
condition for exciting bending waves is $\gamma\la
0.01(\alpha/0.1)$. 

\section{Related work}

\label{sec:other}

\noindent
Most treatments of warped Lense--Thirring discs neglect torques from
the companion in determining the shape and evolution of the disc; we
may call this the Bardeen--Petterson approximation since it
first appears in \cite{bp75}. The approximation is only valid if
the torque associated with viscous angular-momentum transport exceeds
the Lense--Thirring and companion torques at the point where the
latter two are equal, the warp radius $r_w$ (eq.\ \ref{eq:lapk}),
which in turn requires $\beta\ga 1$ (eq.\ \ref{eq:qdef}). 

One of the few treatments of warped AGN accretion discs to include
both Lense--Thirring and tidal torques is \cite{mpt09}. In fact the
warp radius $r_{\rm warp}$ defined in their equation (15) is almost
the same as the radius $r_w$ defined in our equation (\ref{eq:kerr}),
$r_{\rm warp}=r_w/2^{2/9}$. Martin et al.\ also define a tidal radius
$r_{\rm tid}$ and a Lense--Thirring radius $r_{\rm LT}$ where viscous
torques balance tidal and Lense--Thirring torques, respectively. Our
parameter $\beta$, defined in equation (\ref{eq:qdef}), is just
$2^{1/9}(r_{\rm tid}/r_{\rm LT})^{10/9}$. Martin et al.\ find
numerical solutions for steady-state discs with obliquities up to
$80^\circ$ but all their models have $r_{\rm tid}/r_{\rm LT}\ge 1$ and
their models with obliquities $>20^\circ$ have $r_{\rm tid}/r_{\rm
  LT}=10$. Therefore they do not explore the regime with
$\beta\la 1$ where the critical obliquity becomes apparent.

\cite{sf96} give a simple analytic description of warped accretion
discs, derived from the Pringle--Ogilvie equations by linearizing in
the warp angle. The main focus of their analysis is on estimating the
rate at which the BH aligns its angular momentum with that of
the accreting material. Unfortunately, the linearization drops the
term proportional to $|\p\bfn/\p x|^2$ in equation (\ref{eq:cldef}),
and without this term low-viscosity Lense--Thirring discs develop a
thin boundary layer in which the warp angle jumps sharply, so the
linearization is not self-consistent when $\beta$ is sufficiently
small. 

\cite{nk12} and \cite{nix12} have argued that warped discs described
by the Pringle--Ogilvie equations can `break' or `tear' -- divide
into inner and outer parts with discontinuous orientations -- if the
obliquity $\ga 45^\circ$. As described in their papers, this phenomenon
does not appear to be directly related to our critical obliquity, for several reasons:
(i) Nixon \& King do not include torques from a companion in their
analysis, i.e., the parameter $\beta$ in equation (\ref{eq:qdef}) is
very large, whereas we find that the critical obliquity is 
important only for $\beta\la 1$ (Fig.\ \ref{fig:six}). (ii) Nixon \&
King argue that the breaking phenomenon arises through the dependence
of the viscosity parameters $Q_i$ on the warp $\psi$, whereas we have
found that the critical obliquity is almost the same whether or not
this dependence is included in the differential equations. (iii) We do not
see breaks in our high-viscosity ($\beta=1000$)  solutions, even for obliquities
exceeding $88^\circ$, probably because our expression
for $Q_2(\psi)$ is relatively flat (Fig.\ \ref{fig:two}) whereas Nixon
\& King's falls sharply toward zero for $\psi\ga 1$ (their Fig.\
1)\footnote{The reason for this difference has been pointed out to us by G.~Ogilvie
  (private communication). In a flat isothermal disc the sound speed
  and rms thickness are related by $c_s=H\Omega$; however, this
  relation no longer holds in a warped disc because a vertical
  oscillation is present, so hydrostatic equilibrium does not apply. Nixon \& King's
  `isothermal' disc has $H$ independent of the warp angle $\psi$
  whereas ours has $c_s$ independent of $\psi$.}.

\section{Application to observed accretion discs}

\label{sec:observations}

\noindent
The accreting BHs found in astrophysical systems span a wide range of
inferred mass, from $\mbh \sim 5\msun$ up to $\sim 10^{10} \msun$.
Within this range they mostly fall -- so far -- into one of two distinct
classes. At the low-mass end, $\mbh\sim 10 \msun$, the BHs all belong
to close binary systems. The BH accretes mass from its companion star,
either by Roche-lobe overflow or by capturing a fraction of the mass
lost in a wind.  Roche-lobe overflow tends to occur in low mass X-ray
binaries (LMXBs), in which the companion is an evolved star with
$\mstar \la 1.5\msun$.  Wind-driven accretion is found in high
mass X-ray binaries (HMXBs), where the companion is an O or B star
with $\mstar\ga 10\msun$.  The secondary star provides the tidal
torque in equation (\ref{eq:comp}), which is also thought to set the
outer radius of the accretion disc. The dynamics and geometry of
accretion in these systems is relatively well-understood and useful
summaries are found in \citet{2002apa..book.....F} and
\cite{2006ARA&A..44...49R}.

The second class consists of supermassive BHs, with $\mbh\sim
10^5$--$10^{10} \msun$, which are found -- so far -- at the centres of
galaxies and primarily accrete gas from the interstellar medium of
their galaxy.  When mass is supplied at sufficiently high rates, these
are observed as AGN \citep{1999agnc.book.....K}.  The properties of
these systems and how they are fed from the interstellar medium are less well
understood than binary systems and there are fewer empirical
constraints on the properties of the disc\footnote{We do not consider
  the ultraluminous X-ray sources with $L \ga 10^{40}\mbox{\,erg
    s}^{-1}$. If these radiate isotropically and do not exceed the
  Eddington limit, they require BHs with $\mbh\ga 100 \msun$.  Whether
  or not these are, in fact, intermediate-mass BHs or normal HMXBs,
  the implied accretion rates suggest that ultraluminous X-ray sources
  arise from a short-lived phase of rapid mass transfer in a close
  binary \citep{2001ApJ...552L.109K}.}.

We discuss these two classes of Lense--Thirring discs in the
next two subsections. 

\subsection{Stellar-mass black holes in binary star systems}

\label{sec:xrb}

\noindent
In these binaries the X-ray emission comes from the vicinity of a
neutron star or BH (the `primary'), while the accreted mass
and the tidal torque (\ref{eq:comp}) comes from the companion star
(the `secondary'). The masses of the primary and secondary, $M$ and
$\mstar$, and their orbital separation $r_\star$ are inferred from
the orbital period, the spectral type and velocity semi-amplitude of
the secondary, periodic variations in the flux from the secondary due
to its tidal distortion by the primary, eclipses, etc. In most cases
the main evidence that the primary is a BH rather than a neutron
star is that its mass exceeds the upper limit to the mass of a neutron
star, $\sim 3\msun$ \citep{lat05}.

Compilations of BH X-ray binary system parameters can be found in
Tables 4.1 and 4.2 of \citet{2006csxs.book..157M} and Table 1 of 
\citet{2006ARA&A..44...49R}.  The inferred BH masses have a
relatively narrow distribution -- the best estimates in $\sim 20$
systems range from 4.5 to $14\msun$ -- with a mean near $\mbh \sim 7
\msun$.  The BH spin $a_\bullet$ is more difficult to measure.  The
two most commonly used methods are continuum fitting
\citep[e.g.][]{2011CQGra..28k4009M} and Fe line modeling
\citep{1995Natur.375..659T}.  Only a range of plausible spins can be
inferred, even for the best systems, and both methods are subject to
systematic uncertainties.  For our purposes, the most important result
is that the majority of systems are not consistent with $a_\bullet=0$,
implying that Lense--Thirring precession can be significant.  Since the
parameter $\beta$ (eq.\ \ref{eq:qdef}) depends relatively weakly on
$a_\bullet$ ($\beta \propto a_\bullet^{-4/9}$), we simply adopt
$a_\bullet=0.5$ as a characteristic value.

There is strong circumstantial evidence for warps in several X-ray
binaries. The jets in the eclipsing X-ray binary SS 433 precess with a
162 d period, likely because the jet direction is normal to a
precessing warped accretion disc. The 35 d period of Her X-1 is
believed to be due to eclipses by a warped disc, and this is also
the likely explanation for some of the long-term periodicities
observed in other X-ray binaries, such as LMC X-4 and SMC X-1
\citep{cha08}. There is also evidence for misalignment between the
binary orbital angular momentum and BH spin angular momentum in GRO
J1655$-40$ and V4641 Sgr, if one assumes that the jet axis is aligned
with the BH spin axis \citep[e.g.,][]{fmw01,mac02}.

Most BH candidates with mass estimates are LMXBs, and only a handful
are HMXBs. In the Roche-lobe overflow systems that comprise the bulk
of LMXBs, it is thought that the tidal torque from the companion
truncates the accretion disc at an outer radius $r_{\rm out} \simeq
0.9 r_{L1}$ where $r_{L1}$ is the Roche radius\footnote{`Roche radius'
  is defined as the radius of a sphere with the same volume as the
  Roche lobe; the distance to the collinear Lagrange point from the
  centre of the star is larger by $\sim 25$--$40$ per cent, depending on the
  mass ratio. An analytic approximation to the Roche radius as a
  function of mass ratio is given by \citet{1983ApJ...268..368E}.} of
the primary \citep{2002apa..book.....F}.  Fitting of ellipsoidal
variations of LMXBs with BH primaries generally yields $r_{\rm out}$
values consistent with this assumption (J. Orosz, private
communication).

In LMXB systems, the secondaries are generally evolved F-K spectral types
with $\mstar \sim \msun$, so we scale the companion mass $\mstar$ to
$\msun$. Orbital periods $P$ range from a few hours to several days so
we scale the period to $10^5\mbox{\,s}=27.8\mbox{\,h}$.  Then the
separation or semimajor axis is\comment{actual value 9.261}
\begin{equation}
\rstar=\left(\frac{P}{2\upi}\right)^{2/3}[G(\mbh+\mstar)]^{1/3}=9.3\,\rsun\left(\frac{P}{10^5\mbox{\,s}}\right)^{2/3}\left(\frac{\mbh+\mstar}{8\msun}\right)^{1/3}.
\label{eq:routlmxb}
\end{equation}
The large range of $P$ translates into a fairly broad range in $\rstar$.  At the lower end of the
range, corresponding to periods of a few hours, we expect
$\rstar\simeq 2$--$3\rsun$, although $\rstar$ can be much larger than this estimate in
some cases such as GRS 1915+105: here $P=804\mbox{\,h}$ so
$\rstar=87\rsun$ for $\mbh+\mstar=8\msun$. 

For comparison, the warp radius (\ref{eq:kerr}) is \comment{0.1871}
\begin{equation} 
r_w = 0.19\, \rsun
\left(\frac{a_\ast}{0.5}\right)^{2/9}
\left(\frac{\mbh}{7\msun}\right)^{5/9}
\left(\frac{\msun}{\mstar}\right)^{2/9}
\left(\frac{\rstar}{10\,\rsun}\right)^{2/3} \label{eq:rwlmxb}.
\end{equation} 
Assuming a mass ratio $\mbh/\mstar=7$ the primary's Roche radius is
$r_{L1}=0.55\rstar$, so if the outer disc edge is at $r_{\rm out}
\simeq 0.9 r_{\rm L1}$ we have $r_{\rm out} \simeq 0.5\rstar$.  Hence,
for typical LMXBs the warp radius (\ref{eq:rwlmxb}) is well inside the
outer disc radius (cf.\ eq.\ \ref{eq:routlmxb}).

Similar conclusions hold for HMXBs.  We consider the specific example
of M33 X-7 since it is the best-understood HMXB system due to its
X-ray eclipses and well-determined distance
\citep{2007Natur.449..872O,2008ApJ...679L..37L}.  In this case we have
$\mstar = 70\pm7 \msun$, $\mbh=15.7\pm1.5\msun$, $\rstar=42\pm 2
\rsun$, $a_\bullet=0.84\pm0.05$, yielding a warp radius
$r_w=0.34\rsun$. Orosz et al.\ also find that the outer radius of the
disc is $r_{\rm out}=(0.45\pm0.04)r_{L1}$; for the observed mass ratio
$r_{L1}=0.5\rstar$ \citep{1983ApJ...268..368E} so $r_{\rm
  out}=9.5\rsun$. Again, the warp radius is well inside the outer disc
radius\footnote{Note that the common assumption that $r_{\rm
    out}=0.9r_{L1}$ is not confirmed in M33 X-7, where the eclipse
  models give a result a factor of two smaller. In wind-fed HMXBs the
  disc could plausibly be truncated at smaller radii via interactions
  with the wind. Direct constraints on $r_{\rm out}$ in other HMXBs
  are hampered by the dominance of the secondary in the optical band
  \citep[see e.g.][]{2009ApJ...697..573O}.}.

The strength of the viscous torque can be parametrized through the
disc aspect ratio $H/r$, which is related to the sound speed through
$c_s=\Omega H$.  The aspect ratio can be estimated using the standard
thin-disc model of \citet{1973A&A....24..337S}.  In BH X-ray binaries,
the warp radius is much larger than the BH event horizon, so we can
ignore relativistic effects and corrections due to the inner
boundary condition; moreover at the warp radius the radiation pressure
is negligible. We can therefore use equation (\ref{eq:siggas})
below\footnote{Equation 2.16 of \citealt{1973A&A....24..337S} gives
  the same result to within 30 per cent for their assumed efficiency $\epsilon=0.06$.} to
estimate\comment{I get 9.21, Shane gets 9.06}
\begin{equation}
\bigg(\frac{H}{r}\bigg)^2 \simeq 9.1\times10^{-5}
\bigg(\frac{L}{0.01L_{\rm Edd}}\frac{0.1}{\epsilon}\bigg)^{2/5}
\bigg(\frac{0.1}{\alpha}\bigg)^{1/5}
\bigg(\frac{7 \msun}{\mbh}\bigg)^{3/10}
\bigg(\frac{r}{\rsun}\bigg)^{1/10}. \label{eq:hrss}
\end{equation}
We assume that the Shakura--Sunyaev parameter $\alpha$ (eq.\
\ref{eq:alpha}) is approximately 0.1, based on modeling of dwarf novae and soft
X-ray transients \citep{2007MNRAS.376.1740K}. 

This equation is determined by balancing local viscous heating with
radiative cooling.  However, the spectra from the outer regions of
discs in LMXBs show evidence that irradiation by X-rays dominates over
local dissipation \citep{1994A&A...290..133V}.  Simple models of the
X-ray irradiated outer disc imply only a weak dependence of $H/R$ on
$R$ \citep[e.g.,][]{1999MNRAS.303..139D}.  So we make an alternative
estimate of the aspect ratio, valid for the outer parts of the disc,
by scaling to a characteristic temperature $T$ and assuming
hydrostatic equilibrium.  Then we have approximately 
\begin{equation}
\left(\frac{H}{r}\right)^2 \simeq \frac{k T r}{G \mbh m_p}
\simeq 2\times 10^{-4} \frac{r}{3\rsun}
\frac{7\msun}{\mbh}\frac{T}{10^4 \mbox{\,K}}.
\label{eq:hrir}
\end{equation}
Soft X-ray transient LMXBs are believed to be triggered by a disc
instability associated with hydrogen ionization
\citep{2001NewAR..45..449L} so one expects the outer disc has $T
\la 10^4 \mbox{\,K}$ at the beginning of an outburst, but the temperature
may rise to as high as $T \sim 10^5 \mbox{\,K}$ during outburst.

Taken together equations (\ref{eq:hrss}) and (\ref{eq:hrir}) imply $(H/R)^2
\simeq 10^{-5}$--$10^{-3}$ in most discs.  Inserting the above estimates into
equation (\ref{eq:qdef})  we find \comment{122.26}
\begin{equation} 
\beta =120\bigg(\frac{0.5}{a_\bullet}\bigg)^{2/3}
\bigg(\frac{\msun}{\mstar}\bigg)^{1/3}
\bigg(\frac{7\msun}{\mbh}\bigg)^{2/3} \frac{\rstar}{10 
  \rsun}\frac{(H/r)^2}{10^{-4}}
\label{eq:betaxrb}
\end{equation}
where $H/r$ is evaluated at the warp radius. 

Therefore, we generally expect $\beta \gg 1$, that is, viscous torques
are more important than the torque from the secondary star in
determining the warp shape. In order to have the companion torque
dominate the warp dynamics, we need $\alpha_\perp\beta \la 1$, which requires
a nearby companion (the shortest orbital periods of X-ray binaries are
a few hours, corresponding to $r_\star\sim3\rsun$) and, more
importantly, a cool disc with $H/r \la 10^{-3}$. This is
plausible for quiescent discs, with low accretion rates, as
long as irradiation by the central X-ray source does not enforce a
larger $H/r$ at the radius of the warp. One might even speculate that
the absence of a steady-state solution for warped discs with
$\beta\la 1$ is the process that drives disc instability and
outbursts in some X-ray binaries. 

\subsection{Warped discs in active galactic nuclei}

\label{sec:agn}

\noindent
There is strong circumstantial evidence that warps are common in
AGN accretion discs. Maser discs having modest warps on 0.1--1 pc
scales are present in NGC 4258 \citep{her05}, Circinus
\citep{green03}, and four of the seven galaxies examined by
\cite{kuo11}. Warped discs may obscure some AGN and thus play a role
in unification models of AGN based on orientation \citep{nay05}. The
angular-momentum axis of material accreting onto the AGN, as traced by
jets or other indicators, is not aligned with the axis of the host
galaxy on large scales \citep{kin00}. Radio jets from AGN often show
wiggles or bends that may arise from precession of the jet source (e.g., 3C
31). Finally, frequent and variable misalignments of the BH spin axis
with the angular momentum of accreted gas are expected theoretically
because of clumpy gas accretion, inspiral of additional BHs, and rapid
angular-momentum transport within gravitationally unstable gas discs
\citep{hop12}.

AGN accretion discs are much less well-understood than X-ray binary
discs. There is no obvious source of external torque analogous to the
companion star in X-ray binaries -- except in the case of binary BHs,
which we defer to \S\ref{sec:bbh}. In the
absence of external torques, warping can arise from a misalignment
between the orbital angular momentum of the inflowing material at the
outer edge of the disc and the spin angular momentum of the BH at its
centre. Then in the absence of other torques the shape of the
warp is determined by the competition between viscous torques and the
Lense--Thirring torque (the Bardeen--Petterson approximation).

However, AGN discs are much more massive than X-ray binary discs
relative to their host BH, and this raises the possibility that the
self-gravity of AGN discs plays a prominent role in determining the
shape of the disc.

Self-gravitating\footnote{As described in the Introduction, by
  `self-gravitating' we mean that the self-gravity of the warped disc
  dominates the angular-momentum precession rate, not that the disc is
  gravitationally unstable or that its mass is comparable to the BH
  mass.}  warped discs have mostly been investigated in the context of
galaxy discs, which are sometimes warped in their outer parts. There
is a large literature on the dynamics of galactic warps \citep[e.g.,][]{ht69,sc88,jjb92,nt96,
sell13}. Very few authors have examined the properties of
self-gravitating warped discs in the context of AGN. One notable
exception is \cite{ger09}, who computed the shapes of warped
self-gravitating discs orbiting a central mass, modeling the disc as a
set of concentric circular rings and computing the gravitational
torques between each ring pair. However, they did not include either
Lense--Thirring or viscous torques so their calculations do
not address the issues that are the focus of the present paper. 

We first describe a simple analytic model for flat AGN accretion
discs, which we shall use to estimate the relative importance of
self-gravity and viscous stresses in warped discs. Our model is
similar to earlier analytic models by \cite{1973A&A....24..337S},
\cite{pri81}, \cite{cs90}, and others.

We assume that the density $\rho(r,z)$ in the disc is small compared
to $\mbh/r^3$. Then hydrostatic equilibrium requires
\begin{equation}
\frac{d p_t}{dz}=\Omega^2 \gothr_z \rho z,\label{eq:he}
\end{equation}
where $p_t=p_g + p_r$ is the sum of the gas and radiation pressure,
$\Omega^2=G\mbh/r^3$, and $\gothr_z$ is a dimensionless factor
discussed below.  The equation of energy conservation is
\begin{equation}
F_r =\frac{3}{4}\Omega \frac{\gothr_R}{\gothr_T} \int dz\, \tau_{r\phi},\label{eq:ec}
\end{equation}
where $F_r$ is the emissivity from one surface of the disc and
$\tau_{r\phi}$ is the viscous stress tensor.  Together with $\gothr_z$
above, $\gothr_R$ and $\gothr_T$ are dimensionless factors that depend
on radius and the BH spin parameter $a_\bullet$ and approach unity for
$r\gg R_g$, where as usual $R_g=G\mbh/c^2$ is the gravitational radius
of the BH. These quantities, defined in Chapter 7 of
\citet{1999agnc.book.....K}, account approximately for
general-relativistic effects and incorporate the assumption of no
torque at the radius $r_{\rm ISCO}$ of the innermost stable circular
orbit.

Coupling equation (\ref{eq:ec}) to the equation for conservation
of angular momentum in a flat steady-state disc allows one to solve for $F_r$,
\begin{equation}
F_r=\frac{3c^3 (L/L_{\rm Edd})}{2 \kappa R_g \epsilon (r/R_g)^3}
\gothr_R,
\label{eq:flux}
\end{equation}
where $L/L_{\rm Edd}$ is the ratio of the bolometric luminosity of the
disc to the Eddington luminosity, $\kappa$ is the electron scattering
opacity (assumed to be $\simeq 0.34\,\cm^2\mbox{ g}^{-1}$),  and 
$\epsilon=L/(\dot M_\bullet c^2)$ is the radiative efficiency.

We now make the standard $\alpha$-disc approximations that the
stress has the form (eq.\ \ref{eq:alpha}) 
\begin{equation}
\tau_{r\phi}=-\eta r\frac{d\Omega}{dr}=\frac{3}{2}\alpha p_t,
\end{equation}
and that the rate of energy dissipation per unit mass is independent
of $z$. Then the radiation pressure and the temperature at the midplane of the disc are
\begin{equation}
p_{r0}=\frac{F_r \kappa \Sigma}{4 c}, \quad T_0=\left(\frac{3F_r\kappa}{16\sigma_B}\right)^{1/4},
\end{equation}
where $\sigma_B$ is the Stefan-Boltzmann constant.
The gas pressure at the midplane is 
\begin{equation}
p_{g0}=\frac{\rho_0k_BT_0}{\mu}=\frac{\rho_0k_B}{\mu }\left(\frac{3 F_r \kappa}{16 \sigma_B}\right)^{1/4}
=\left(\frac{3 F_r \kappa}{16 \sigma_B}\right)^{1/4}\frac{k_B}{\mu }\frac{\Sigma^{5/4}}{2H},
\end{equation}
where $k_B$ and $\rho_0$ are Boltzmann's constant and the
midplane density. The mean particle mass $\mu$ is taken to be the
proton mass times 0.62, appropriate for fully ionized hydrogen plus
30 per cent helium by mass. In the last equation
we have replaced $\rho_0$ by $\Sigma/(2H)$ where $H$ is the disc
thickness. 

We now substitute these results into equations (\ref{eq:he}) and (\ref{eq:ec}) with
the replacements $d/dz \rightarrow 1/H$, $z \rightarrow H$, and $\int\,dz \rightarrow 2H$,
to obtain
\begin{equation}
\frac{F_r \kappa \Sigma}{4 c} +
\left(\frac{3 F_r \kappa}{16 \sigma_B}\right)^{1/4}\frac{k_B}{\mu }
\frac{\Sigma^{5/4}}{2H}-\frac{\Omega^2 \gothr_z H \Sigma}{2}=0\label{eq:he1z}
\end{equation}
and
\begin{equation}
\frac{H F_r \kappa \Sigma}{4 c} +
\left(\frac{3 F_r \kappa}{16 \sigma_B}\right)^{1/4}\frac{k_B}{2 \mu }
\Sigma^{5/4}
-\frac{4 F_r \gothr_T}{9 \Omega \alpha \gothr_R }=0.\label{eq:ec1z}
\end{equation}
For given values of the radius $r$, the gravitational radius $R_g$,
the efficiency $\epsilon$, and the Eddington ratio $L/L_{\rm Edd}$,
the second of these equations can be solved for the disc thickness
$H$. Then the result can be substituted into the first equation to
yield a tenth degree polynomial in $\Sigma^{1/4}$, which can be solved
numerically to find the surface density \citep{2012MNRAS.424.2504Z}.

The analysis is simpler when the accretion disc is dominated by
radiation pressure or gas pressure. For radiation-pressure dominated
discs we set $p_g=0$ in equations  (\ref{eq:he1z}) and (\ref{eq:ec1z}).
We then find\comment{69.7, 11.08}
\begin{align}
\Sigma_r &=\frac{2^6}{3^3}\,\frac{\gothr_z \gothr_T}{\gothr_R^2}\,\frac{\epsilon}{\alpha\kappa}\,\frac{L_{\rm
    Edd}}{L}\left(\frac{r}{R_g}\right)^{3/2}= 70\mbox{\;g cm}^{-2}\;
\frac{\gothr_z\gothr_T}{\gothr_R^2}\;\frac{\epsilon}{0.1}\;\frac{0.1}{\alpha}\;\frac{0.1L_{\rm
    Edd}}{L}\left(\frac{r}{R_g}\right)^{3/2}\nonumber \\[5pt]
H_r &=  \frac{3 \gothr_R}{4\gothr_z}\;\frac{L}{L_{\rm Edd}}\;\frac{R_g}{\epsilon}= 1.1\times 10^{13}\cm\,
\frac{\gothr_R}{\gothr_z}\;\frac{0.1}{\epsilon}\;\frac{L}{0.1L_{\rm Edd}}\;\frac{\mbh}{10^8\msun}.
\label{eq:sigrad}
\end{align}

Similarly, when radiation pressure is negligible, \comment{1.402,2.469}
\begin{align}
\Sigma_g &=\frac{2^{14/5}\upi}{3^{7/5}5^{1/5}}\;\frac{\mu^{4/5}(G\mbh
  c)^{1/5}}{\kappa^{4/5} h^{3/5}\alpha^{4/5}\epsilon^{3/5}}
\;\frac{\gothr_T^{4/5}}{\gothr_R^{1/5}} \left(\frac{L}{L_{\rm Edd}}\;\frac{R_g}{r}\right)^{3/5}
\nonumber \\[5pt]
&= 1.4\times10^7\; {\rm g \; cm^{-2}} \;
\frac{\gothr_T^{4/5}}{\gothr_R^{1/5}}\left(\frac{0.1}{\alpha}\right)^{4/5}\left(\frac{\mbh}{10^8\msun}\right)^{1/5}\left(\frac{0.1}{\epsilon}\;\frac{L}{0.1L_{\rm Edd}}\;\frac{R_g}{r}\right)^{3/5}
\label{eq:siggas}
\end{align}
and
\begin{align}
H_g &=\frac{3^{1/5}5^{1/10}}{2^{2/5}\upi^{1/2}}\;\frac{h^{3/10}(G\mbh)^{9/10}
}{\mu^{2/5}\kappa^{1/10}c^{21/10}\alpha^{1/10}\epsilon^{1/5}}\;\frac{\gothr_R^{1/10}\gothr_T^{1/10}}{\gothr_z^{1/2}}\left(\frac{L}{L_{\rm
      Edd}}\right)^{1/5}\left(\frac{r}{R_g}\right)^{21/20} \nonumber \\[10pt]
&= 2.5\times 10^{10}\;\cm\; \frac{\gothr_R^{1/10}\gothr_T^{1/10}}{\gothr_z^{1/2}}\left(\frac{0.1}{\alpha}\right)^{1/10}\!\left(\frac{\mbh}{10^8\msun}\right)^{9/10}\!\left(\frac{0.1}{\epsilon}\right)^{1/5}\!\left(\frac{L}{0.1L_{\rm Edd}}\right)^{1/5}\!\left(\frac{r}{R_g}\right)^{21/20}\!\!.
\end{align}

With these scalings, we can compute most properties of interest
in the disc.  For example, radiation pressure dominates when $H_r>H_g$
which occurs for radii less than\comment{4.955}
\begin{equation}
r_{pr} \simeq 5.0\times10^{15}\;\cm\;
\left(\frac{\alpha}{0.1}\right)^{2/21}
\left(\frac{0.1}{\epsilon}\;\frac{L}{0.1L_{\rm
      Edd}}\right)^{16/21}\left(\frac{\mbh}{10^8\msun}\right)^{23/21}
\;\frac{\gothr_R^{6/7}}{\gothr_z^{10/21}\gothr_T^{2/21}}\bigg|_{r_{pr}}.\label{eq:rpr}
\end{equation}
The disc is gravitationally unstable if Toomre's (1964) $Q$ parameter
is less than unity; this parameter is approximately
\begin{equation}
Q=\frac{\Omega^2 H}{\upi G \Sigma}.
\label{eq:toomredef}
\end{equation}
In the radiation- and gas-pressure dominated regimes
(respectively) we have \comment{3.124,3.459}
\begin{align}
Q_r &= 3.1 \times 10^{12} \;\frac{\gothr_R^3}{\gothr_z^2\gothr_T}
\left(\frac{L}{0.1L_{\rm Edd}}\;\frac{0.1}{\epsilon}\right)^2\;\frac{10^8\msun}{\mbh}\;\frac{\alpha}{0.1}\left(\frac{R_g}{r}\right)^{9/2} \nonumber \\[10pt]
Q_g &= 3.5\times 10^{4}\; \frac{\gothr_R^{3/10}}{\gothr_T^{7/10}\gothr_z^{1/2}}
\left(\frac{0.1L_{\rm
      Edd}}{L}\;\frac{\epsilon}{0.1}\right)^{2/5}\left(\frac{10^8\msun}{\mbh}\right)^{13/10}\left(\frac{\alpha}{0.1}\right)^{7/10}\left(\frac{R_g}{r}\right)^{27/20}.
\label{eq:toomre}
\end{align}

Similarly, we can compute the warp radius (eq.\ \ref{eq:rwself})\comment{4.296e15,3.874e15}
\begin{align}
r_{w,r} &= 4.3\times10^{15}\;\cm\;\left(\frac{a_\bullet}{0.5}\right)^{1/5}
\left(\frac{\alpha}{0.1}\;\frac{0.1}{\epsilon}\;\frac{L}{0.1L_{\rm
      Edd}}\right)^{1/5}\left(\frac{\mbh}{10^8\msun}\right)^{4/5}
\;\frac{\gothr_R^{2/5}}{\gothr_T^{1/5}\gothr_z^{1/5}}\bigg|_{r_{w,r}}\nonumber
\\[10pt]
r_{w,g} &= 3.9\times10^{15}\,\cm\,
\left(\frac{a_\bullet}{0.5}\right)^{10/29}\!\!\left(\frac{\alpha}{0.1}
\right)^{8/29}\!\!\left(\frac{\epsilon}{0.1}\right)^{6/29}\!\!\left(\frac{0.1L_{\rm
  Edd}}{L}\right)^{6/29}\!\!\left(\frac{\mbh}{10^8\msun}\right)^{17/29}\!\!
\frac{\gothr_R^{2/29}}{\gothr_T^{8/29}}\bigg|_{r_{w,g}}.
\label{eq:warprad}
\end{align}
Equation (\ref{eq:warprad}) gives implicit relations for $r_w$ because
of the radial dependence of the relativistic factors.  However, this
dependence is rather weak for typical AGN disc models: for the case
$a_\bullet=0.5$, $M=10^8 \msun$, $\alpha=0.1$ and $L/L_{\rm
  Edd}=0.1$, we have $\gothr_R=0.81$, $\gothr_T=0.81$, and
$\gothr_z=1.01$ at $r_w$, corresponding to values of 0.96 and 1.05 for
the products of relativistic factors in the radiation-pressure and
gas-pressure dominated limits of equation (\ref{eq:warprad}).

The characteristic ratio of the viscous and self-gravity torques is
(cf.\ eq.\ \ref{eq:betaself})\comment{0.1410,0.0563}
\begin{align}
\gamma&=\frac{c_s^2}{\upi G\Sigma r}\bigg|_{r_w}=\frac{H^2
  \Omega^2}{\upi G \Sigma r}\bigg|_{r_w}\nonumber \\[5pt]
&=0.14\left(\frac{0.5}{a_\bullet}\right)^{11/10}
\left(\frac{0.1}{\alpha}\right)^{1/10}
\left(\frac{0.1}{\epsilon}\;\frac{L}{0.1L_{\rm
      Edd}}\right)^{19/10}\left(\frac{\mbh}{10^8\msun}\right)^{1/10}
\;\frac{\gothr_R^{9/5}\gothr_T^{1/10}}{\gothr_z^{19/10}}\bigg|_{r_{w,r}}
\nonumber \\[5pt]
&=0.056\left(\frac{0.5}{a_\bullet}\right)^{13/29}
\left(\frac{\alpha}{0.1}\right)^{7/29}
\left(\frac{0.1}{\epsilon}\;\frac{L}{0.1L_{\rm
      Edd}}\right)^{2/29}\left(\frac{10^8\msun}{\mbh}\right)^{25/29}
\;\frac{\gothr_R^{9/29}}{\gothr_T^{7/29}\gothr_z}\bigg|_{r_{w,g}}
\label{eq:gamxxx}
\end{align}
where as usual the two equations correspond to the radiation-pressure
dominated and the gas-pressure dominated regions.

Thus, in our fiducial case -- a disc surrounding a $10^8\msun$ BH 
radiating at 10 per cent of the Eddington luminosity, with spin parameter
$a_\bullet=0.5$, efficiency $\epsilon=0.1$, and Shakura--Sunyaev
parameter $\alpha=0.1$ -- the gravitational radius is $R_g=1.48\times
10^{13}\cm$; the warp radius is just inside the radiation-pressure
dominated region at $r_w=4.3\times 10^{15}\cm=290R_g$; the disc becomes
gas-pressure dominated outside $r_{pr}=5.0\times10^{15}\cm\simeq 340R_g$; the
disc becomes gravitationally unstable outside
$3.4\times10^{16}\cm\simeq 2300R_g$; and the disc warp
is governed by Lense--Thirring and self-gravitational torques, with
viscous torques smaller by a factor of
$\gamma\alpha_\perp\simeq0.14\alpha_\perp$ where $\alpha_\perp\sim 1$
for a Shakura--Sunyaev parameter $\alpha\simeq0.1$.

We supplement these formula with three sets of plots. These plots are
based on the analysis in equations (\ref{eq:he})--(\ref{eq:ec1z}) with
three refinements to the analytic formulae
(\ref{eq:sigrad})--(\ref{eq:gamxxx}): (i) we include both gas and
radiation pressure at all radii; (ii) we include the effects of the
relativistic parameters $\gothr_z$, $\gothr_T$, and $\gothr_R$; (iii)
we compute the efficiency $\epsilon$ from the spin parameter $a_\bullet$
using the estimates from \citet{nt73}.  Thus the plots assume thin-disc
accretion with no torque at the inner boundary, which is assumed to
lie at $r_{\rm ISCO}$, the radius of the innermost stable circular orbit.

\begin{figure}
\includegraphics[width=0.9\textwidth,bb=31 0 610 465]{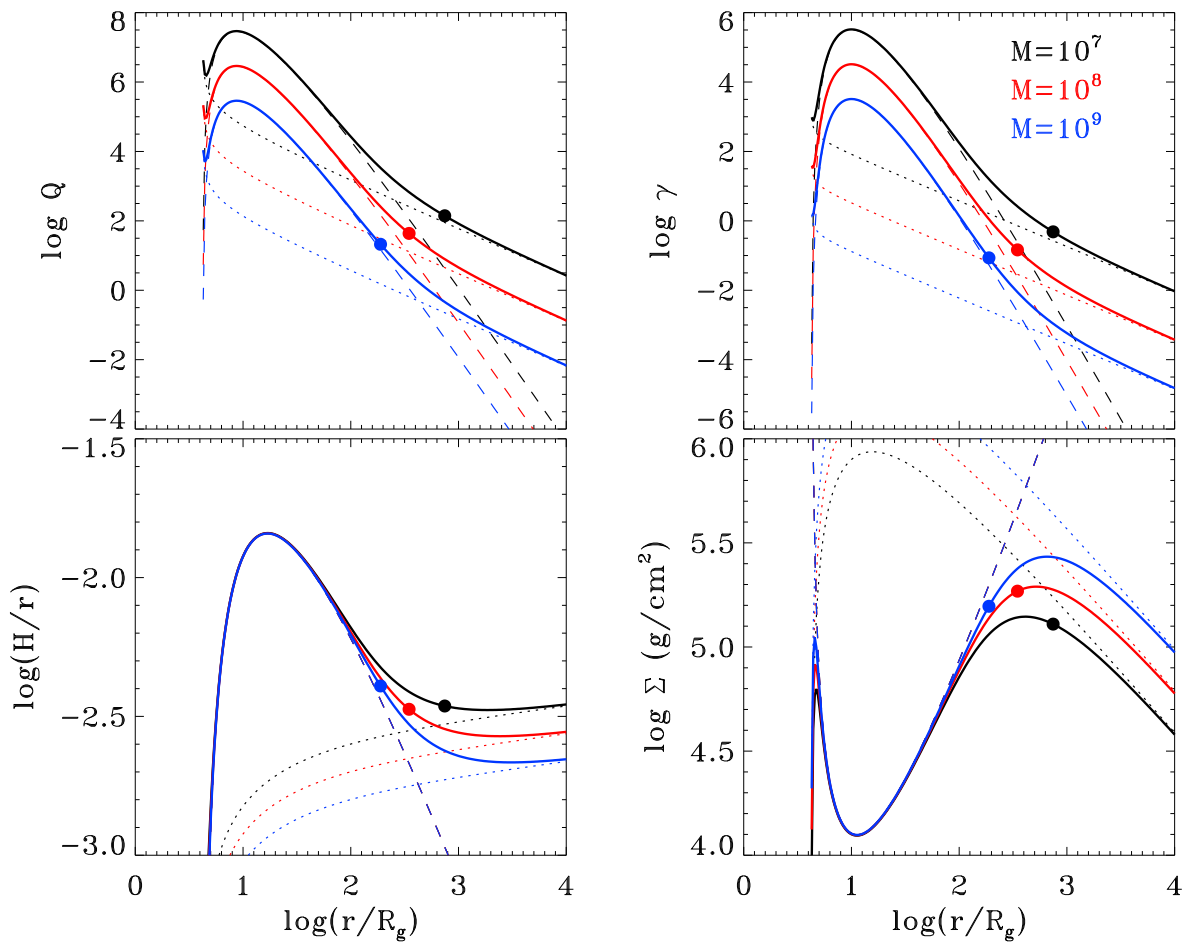}
\caption{Properties of AGN accretion discs with $\alpha=0.1$,
  $a_\bullet=0.5$, $L/L_{\rm Edd}=0.1$, and BH masses $10^7 \msun$
  (black), $10^8 \msun$ (red), and $10^9 \msun$ (blue). The plots
  show Toomre's $Q$ parameter (top left panel), the ratio $\gamma$
  (eq.\ \ref{eq:betaself}) of self-gravity to viscous torque (top
  right), the aspect ratio $H/r$ (bottom left) and the surface density
  (bottom right) versus radius in units of the gravitational radius
  $R_g=G\mbh/c^2$.  The solid curves are computed via direct numerical
  solution of equations (\ref{eq:he1z}) and (\ref{eq:ec1z}), while the
  dashed and dotted curves show the analytic approximations assuming
  that radiation and gas pressure (respectively) dominate. The warp
  radii are marked by filled circles. }
\label{fig:mass}
\end{figure}

\begin{figure}
\includegraphics[width=0.9\textwidth,bb=31 0 610 465]{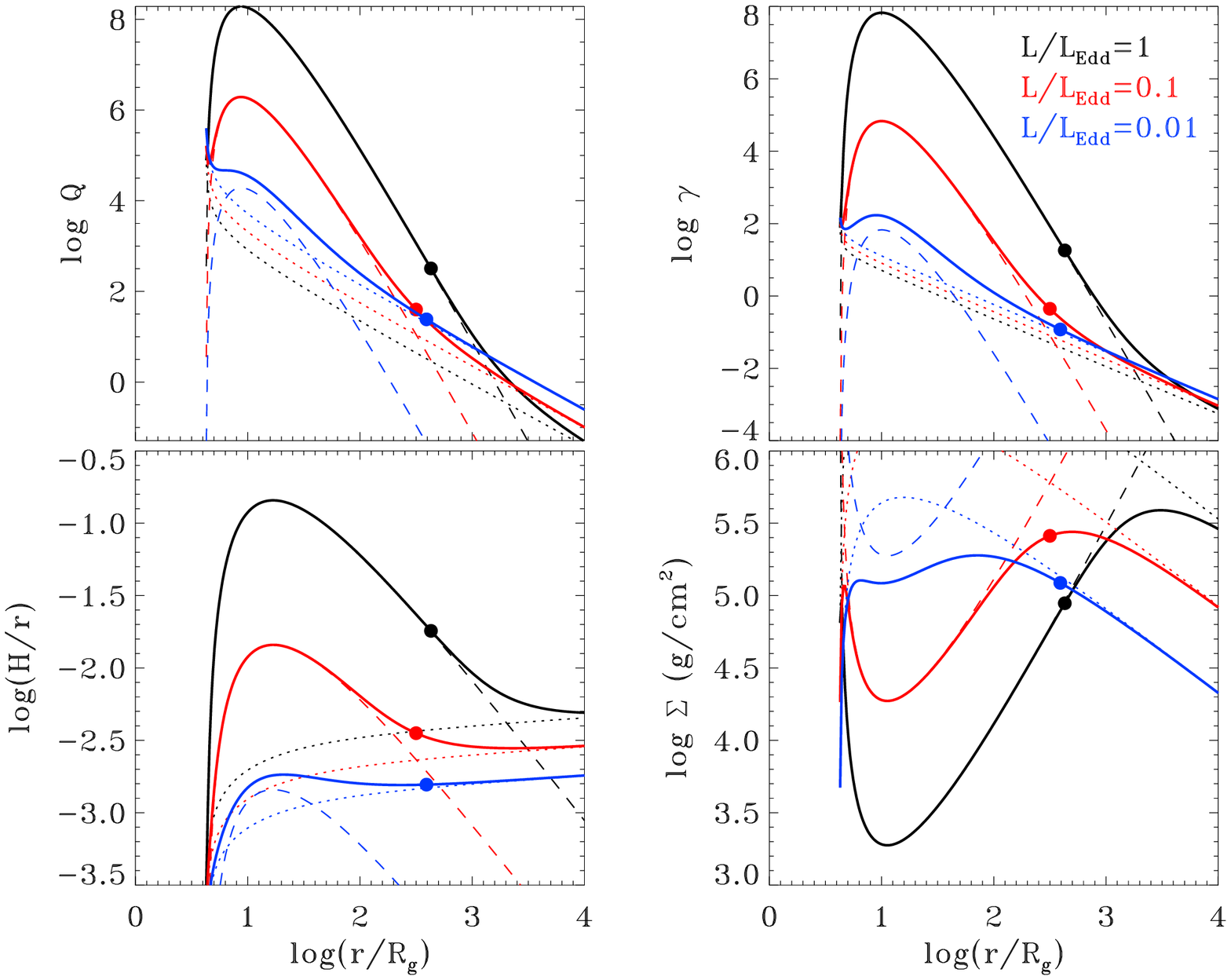}
\caption{As in Fig.\  \ref{fig:mass}, except for BH mass $10^8\msun$
  and Eddington ratios of $1$ (black), $0.1$ (red),
and $0.01$ (blue). }
\label{fig:ledd}
\end{figure}

Fig.\  \ref{fig:mass} shows Toomre's $Q$ (eq.\ \ref{eq:toomredef}),
the aspect radio $H/r$, the surface density $\Sigma$, and the ratio
$\gamma$ of viscous and self-gravity torques for BH masses of
$10^7\msun$, $10^8\msun$, and $10^9\msun$.  Fig.\  \ref{fig:ledd}
shows a similar plot for Eddington ratios $L/L_{\rm Edd}$ of 1, 0.1,
and 0.01.  Figs.\ \ref{fig:mass} and \ref{fig:ledd} show that the
transition from radiation pressure to gas pressure dominance occurs in
the range of 100 to $10^4 R_g$, and depends more strongly on $L/L_{\rm
  Edd}$ than $M_\bullet$.  The radii where $Q$ declines below unity
(onset of local gravitational instability) and $\gamma$ declines below
unity (self-gravity torque stronger than viscous torque) are not very
different, so care must be taken when applying analytic formulae
that assume either radiation or gas pressure to dominate.

\begin{figure}
\includegraphics[width=0.9\textwidth,bb=31 0 610 465]{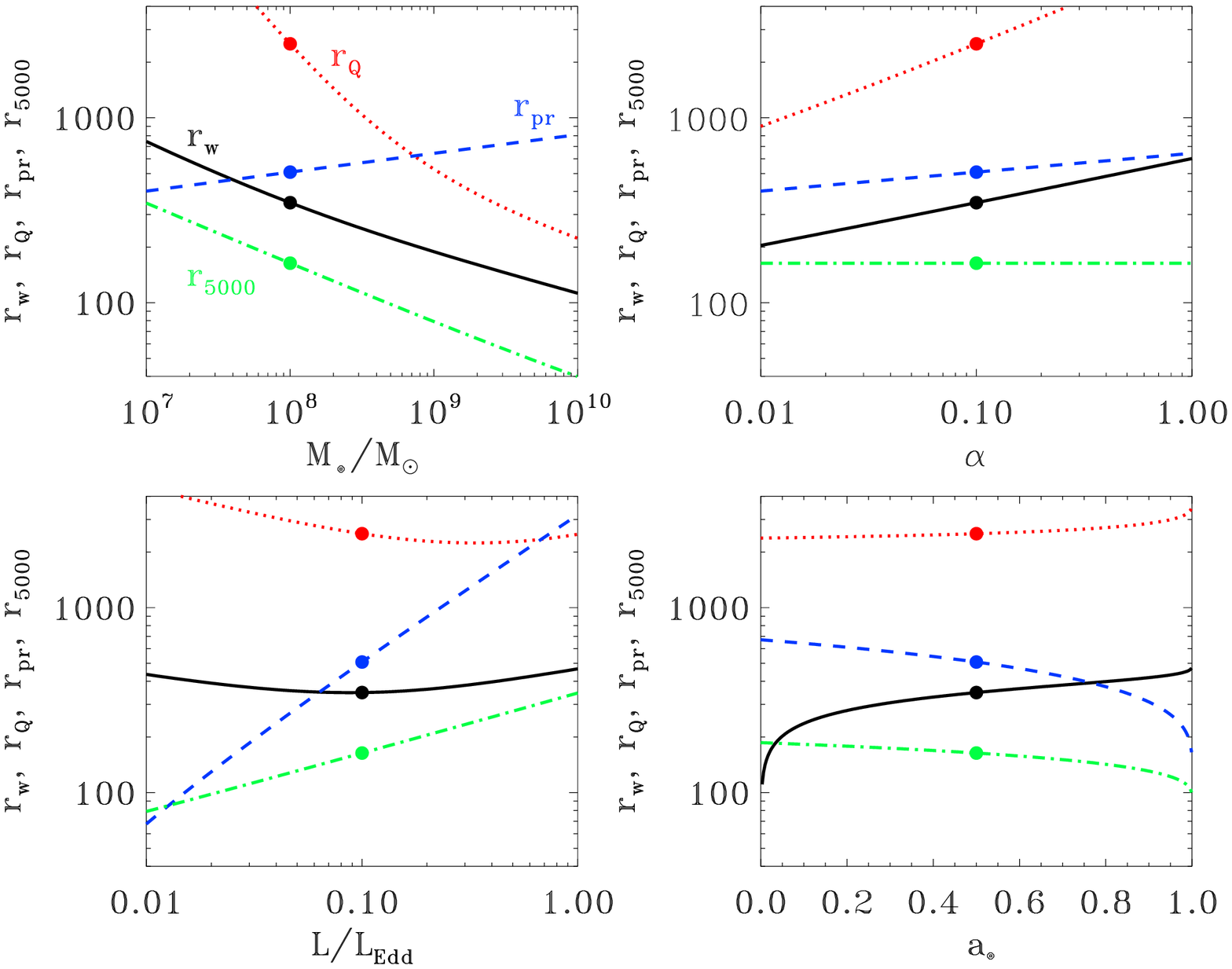}
\caption{Characteristic disc radii versus BH mass (top left panel),
  Shakura--Sunyaev parameter $\alpha$ (top right), Eddington ratio
  (bottom left), and BH spin (bottom right). The curves represent the
  warp radius $r_w$ (eq.\ \ref{eq:rwself}; solid black line), radius
  $r_Q$ at which the disc becomes gravitationally unstable (dotted red
  line), transition radius from radiation-pressure to gas-pressure
  dominated $r_{pr}$ (dashed blue line) and the half-light radius at
  $5000$~\AA (dot-dashed green line).  The fiducial model has
  $M_\bullet =10^8 \msun$, $a_\bullet=0.5$, $L/L_{\rm Edd}=0.1$, and
  $\alpha=0.1$, and is marked by filled circles on each curve. Only a
  single parameter is varied away from the fiducial value to produce
  each panel. All radii are measured in units of the gravitational
  radius $R_g=G\mbh/c^2$.}
\label{fig:comprad}
\end{figure}

Fig.\  \ref{fig:comprad} compares the warp radius $r_w$ to three
characteristic disc radii for a range of disc parameters.  We have
defined the self-gravity radius $r_Q$ as the radius where $Q=1$,
$r_{pr}$ as the radius where the gas and radiation pressure are equal
(cf.\ eq.\ \ref{eq:rpr}), and $r_{5000}$ as the half-light radius for
emission at $5000$~\AA, assuming that the disc radiates locally as a
blackbody.  Since $\gamma$ is smaller than $Q$ by a factor of $ H/r$
(see discussion following eq.\ \ref{eq:betaself}), we always have $r_w
< r_Q$.  The disc is generally in the radiation-dominated regime at
$r_w$, but can fall in the gas-pressure dominated region for smaller
BH mass $M_\bullet$, smaller Eddington ratio $L/L_{\rm Edd}$, or spin
parameter $a_\bullet$ near unity.  The dependence of all the
characteristic radii on $a_\bullet$ is rather weak, except for
$a_\bullet \rightarrow 0$ or 1.

Note that for $\alpha\simeq0.1$ all of the discs shown in these
figures have $\alpha\gg H/r$ (except for $r\la 100 R_g$ when
$L/L_{\rm Edd}=1$) so the condition (\ref{eq:nonres}) for non-resonant
warp behavior is satisfied by a large margin.

For most of the parameter space we have examined the warp radius $r_w$
is just outside (1--3 times larger than) the optical
radius $r_{5000}$. However, if warping causes the disc to intercept a
larger fraction of the emission from smaller radii the region where
the warp is strong may dominate the optical emission. The flux of
radiation coming from the inner disc that irradiates the outer disc is
approximately
\begin{equation}
F_{\rm irr} \approx \frac{L_{\rm in}}{4 \upi r^2} \cos \theta
\end{equation}
where $L_{\rm in}$ is the characteristic luminosity from the inner
disc and $\theta$ is the angle between the normal to the warped outer
disc and the incoming flux.  For thin discs, $\cos \theta \simeq H/r
\ll 1$ and, since $H$ is independent of $r$ in the radiation-dominated
regime, $F_{\rm irr} \propto r^{-3}$.  This is the same scaling as the
intrinsic disc emission (eq.\ \ref{eq:flux}) so disc irradiation has little effect on the
radial emission profile of an unwarped disc.  However, if the disc
has a significant warp, $\cos \theta \gg H/r$ and the irradiating flux
can exceed the intrinsic disc emission.  In this case the
characteristic disc temperature will be
\begin{equation}
T_{\rm irr} \approx  \left(\frac{\chi L}{\upi \sigma_B r_{w,r}^2}\right)^{1/4} 
 \approx  1.1 \times 10^4 \; {\rm K} \;  \left(\frac{\chi}{0.01}\right)^{1/4}
\left(\frac{0.5}{a_\bullet}\;\frac{0.1}{\alpha}\;\frac{\epsilon}{0.1}\right)^{1/10}
\left(\frac{L}{L_{\rm Edd}}\right)^{3/20}\left(\frac{10^8\msun}{\mbh}\right)^{3/20},
\end{equation}
where $\chi$ is a (poorly constrained) reduction factor added to
account for the fraction of the disc luminosity intercepted by the
warp, the characteristic emitting area of the warp, and the albedo.
The wavelength at which blackbody emission peaks for $T_{\rm irr}= 1.1
\times 10^4 \; {\rm K}$ is $\lambda \simeq c h/3 k_B T_{\rm
  irr}=4400$\AA.  Since $r_w$ exceeds the the nominal half-light radius
of the unirradiated disc, the reradiated emission at the
warp can easily dominate.  If so, the true half-light radius for
optical emission should be roughly given by $r_w$ rather than $r_{5000}$.

This result is relevant to recent constraints on the size of quasar
emission regions obtained by modeling the variability due to
gravitational microlensing in an intervening galaxy. In the majority
of cases that have been studied, the sizes inferred from microlensing
exceed the predicted half-light radii of flat $\alpha$-disc models by
factors of $\sim 3$--10 \citep[e.g.][]{mor05,poo07}.  \citet{mor10}
find a best fit in which the microlensing size at $2500$\AA~ scales as
$\mbh^{0.8}$ for a sample of 11 sources with estimated $\mbh=4 \times
10^7 \msun$--$2.4 \times 10^9 \msun$.  This is the same scaling as
$r_{w,r}$ with $\mbh$ in equation (\ref{eq:warprad}) and also agrees
well with the dependence of the warp radius on $\mbh$ found in Fig.\ 
\ref{fig:comprad}. Unfortunately this is not a very sensitive test:
for a flat disc, the radius at a given temperature scales as
$\mbh^{2/3}$, and in the Bardeen--Petterson approximation the warp
radius scales as $\mbh^{9/8}$. The absolute scale for the microlensing
size at 2500\AA~ is a factor of $\sim 6$ smaller than our estimate for
$r_{w,r}$, but this is subject to some uncertainty and might be
accounted for by bending waves excited interior to $r_w$ (compare
Fig.\ \ref{fig:sg}).

An important but poorly understood issue is what fraction of AGN
accretion discs are likely to be warped. Over long times, warps are
damped out as the BH spin axis aligns with the outer disc. A rough
estimate of this time-scale is $t_{\rm align}\simeq L_\bullet/(\upi
r^2\Sigma T_{\rm LT})_{r_w}$ where $L_\bullet$ is the spin angular momentum
of the BH and the quantity in parentheses is the Lense--Thirring
torque per unit mass $T_{LT}$ times the disc mass evaluated at the
warp radius $r_w$. Using equation (\ref{eq:lt}) and the expression for
$L_\bullet$ given just above it, we find
\begin{equation}
t_{\rm align}\simeq
\frac{\mbh}{2\upi  cR_g^{3/2}}\left(\frac{r^{1/2}}{\Sigma}\right)_{r_w}=\frac{r_w^4}{2ca_\bullet R_g^3}.
\end{equation}
where in the second expression we have used (\ref{eq:rwself}) to
eliminate the surface density. For our fiducial
case -- $\mbh=10^8\msun$, $L=0.1L_{\rm Edd}$, $a_\bullet=0.5$,
$\epsilon=0.1$, $\alpha=0.1$ -- the warp radius is $\sim 300R_g$
and\comment{actual number 1.2674e5}
$t_{\rm align}=1.3\times10^5\yr(r_w/300R_g)^4$, much shorter than
the typical AGN lifetime (the Salpeter time, $5\times10^7\mbox{\;yr}$
for $\epsilon=0.1$). Much more uncertain is the time-scale on which
warps are excited. High-resolution simulations of the centres of
galaxies show order unity variations in the gas inflow rate at 0.1 pc
on time-scales less than $10^5\yr$ \citep[][fig.\  6]{hop10} and these are presumably
accompanied by similar variations in the angular momentum of the
inflowing gas. In such an environment the orientation of the outer
parts of the accretion disc is likely to vary stochastically on
time-scales less than the damping time, and this case most AGN
accretion discs will be warped. 

\subsubsection{Binary black holes}

\label{sec:bbh}

\noindent
Most galaxies contain supermassive BHs at their centres, and when
galaxies merge these BHs will spiral to within a few parsecs of the
centre of the merged galaxy through dynamical friction
\citep[e.g.,][]{bbr80,yu02}. Whether they continue to spiral to
smaller radii remains unclear, but if the binary decays to a
sufficiently small semimajor axis -- typically 0.1--0.001 pc, depending
on the galaxy and the BH mass ratio -- the loss of orbital energy
through gravitational radiation will ensure that they merge. If one of
the BHs (the primary) supports an accretion disc, and the spin axis of
the primary is misaligned with the orbital axis of the binary, the
accretion disc will be warped\footnote{There can also be a
  circumbinary accretion disc, which may also be warped, but the
  structure of such discs is poorly understood and we will not discuss
  them here.}. In this case both the self-gravity of the disc and the
tidal field from the secondary, as well as viscous stresses and the
Lense--Thirring effect, can play important roles in shaping the
warp. For the sake of simplicity, we do not examine all of these
torques simultaneously: here we first consider an AGN accretion disc
without self-gravity orbiting one member of a binary BH, then compare the strength of the
torques and the characteristic warp radius to those in an accretion
disc with self-gravity orbiting an isolated BH.

Let $\mbh$ be the mass of the primary and $\mu\mbh$ the mass of the
other BH (the secondary). We assume for simplicity that the orbit is
circular, with semimajor axis $\rstar$. The time required for the two
BHs to merge due to gravitational radiation is \citep{pet64}
\begin{equation}
t_{\rm merge}=\frac{5}{256}\frac{c^5\rstar^4}{G^3\mbh^3\mu(1+\mu)}.
\label{eq:peters}
\end{equation}
The numbers and orbital distribution of binary BHs are not
well-constrained, either observationally or theoretically \citep[see,
for example,][]{shen13}. In the absence of other information, a
natural place to prospect for binary BHs is where the merger time
(\ref{eq:peters}) is equal to the Hubble time. Thus we will use
equation (\ref{eq:peters}) to eliminate the unknown semimajor axis
$\rstar$ in favor of the ratio $t_{\rm merge}/10^{10}\yr$. With this
substitution and using the accretion disc models from earlier in this
Section, most properties of interest are straightforward to
calculate. 

The binary semimajor axis is\comment{1.9876}
\begin{equation}
\rstar=2.0\times10^{17}\cm\;[\mu(1+\mu)]^{1/4}\left(\frac{\mbh}{10^8\msun}\right)^{3/4}\left(\frac{t_{\rm
      merge}}{10^{10}\yr}\right)^{1/4}.
\end{equation}
The warp radius (\ref{eq:kerr}) is\comment{8.9078}
\begin{equation}
r_w=
8.9\times10^{15}\cm\;\frac{(1+\mu)^{1/6}}{\mu^{1/18}}\left(\frac{a_\bullet}{0.5}\right)^{2/9}\left(\frac{\mbh}{10^8\msun}\right)^{5/6}
\left(\frac{t_{\rm merge}}{10^{10}\yr}\right)^{1/6}.
\end{equation} 
The viscosity parameter $\beta$ (eq.\ \ref{eq:qdef}) depends on
whether the warp radius is in the radiation-pressure dominated or the
gas-pressure dominated regime. In these two cases:\comment{0.02292,0.0785}
\begin{align}
  \beta_r &=0.023
  \;\frac{\mu^{1/36}}{(1+\mu)^{1/12}}\left(\frac{0.5}{a_\bullet}\right)^{10/9}\left(\frac{L}{0.1L_{\rm
        Edd}}\right)^2\;\left(\frac{0.1}{\epsilon}\right)^2\;\left(\frac{\mbh}{10^8\msun}\right)^{1/12}\left(\frac{10^{10}\yr}{t_{\rm
        merge}}\right)^{1/12}\frac{\gothr_R^2}{\gothr_T^2}\bigg|_{r_{w}}\nonumber \\[15pt]
  \beta_g &=0.079
  \;\frac{(1+\mu)^{4/15}}{\mu^{4/45}}\left(\frac{0.5}{a_\bullet}\right)^{29/45}\left(\frac{0.1}{\alpha}\right)^{1/5}\left(\frac{L}{0.1L_{\rm
        Edd}}\right)^{2/5}\left(\frac{0.1}{\epsilon}\right)^{2/5}\left(\frac{10^8\msun}{\mbh}\right)^{7/15}\left(\frac{t_{\rm merge}}{10^{10}\yr}\right)^{4/15}\frac{\gothr_R^{1/5}\gothr_T^{1/5}}{\gothr_z}\bigg|_{r_{w}}.
\end{align}
For our fiducial case -- $\mbh=10^8\msun$, $L=0.1L_{\rm Edd}$,
$a_\bullet=0.5$, $\epsilon=0.1$, $\alpha=0.1$, $t_{\rm
  merge}=10^{10}\yr$, $\mu=1$ -- the disc becomes gas-pressure
dominated at $\sim 330R_g$ (eq.\ \ref{eq:rpr}), the warp radius is
$\sim 700R_g$, the disc becomes gravitationally unstable at $2300R_g$
(eq.\ \ref{eq:toomre}), the binary semimajor axis is
$1.6\times10^4R_g$, and the viscosity parameter is
$\beta_g=0.094$. For comparison, including self-gravity leads to a
warp radius of $\sim300R_g$ in an isolated disc (see discussion
following eq.\ \ref{eq:gamxxx}), so self-gravity is likely to have a
stronger influence on the warp shape than torques from the companion
BH, at least in the fiducial disc. Companion torques become stronger
relative to self-gravity in binary BHs with shorter merger times
$t_{\rm merge}$; of course, such systems are relatively rare because
they last for less than a Hubble time.

\section{Summary}

\label{sec:summary}

\noindent
Warped accretion discs exhibit a remarkably rich variety of
behavior. This richness arises for several reasons. First, a
number of different physical mechanisms can lead to torques on the disc:
the quadrupole potential from the central body (e.g., an oblate planet
or a binary black hole), Lense--Thirring precession, the self-gravity
of the disc, the tidal field from a companion, angular-momentum
transport by viscous or other internal disc stresses, radiation pressure, and
magnetic fields (we do not consider the latter two effects). Second,
the geometry of the disc depends critically on whether the competing
mechanisms lead to prograde or retrograde precession of the disc angular
momentum around their symmetry axes. Third, a disc can support
short-wavelength bending waves even when the disc mass is much smaller
than the mass of the central body (as in Saturn's rings).

Most previous studies of warped accretion discs around black holes
have focused on Lense--Thirring and viscous torques (the
Bardeen--Petterson approximation). If a companion star is present in
the system, as in X-ray binary stars, the Bardeen--Petterson
approximation is valid (a `high-viscosity' disc) only if the disc
viscosity is sufficiently high, $\beta\alpha_\perp\ga 1$ where
$\beta$ is given in equation (\ref{eq:betaxrb}) for typical X-ray
binary parameters and $\alpha_\perp\sim 1$ is the Shakura--Sunyaev
$\alpha$ parameter for the internal disc stresses that damp the
warp. Our results suggest that the Bardeen--Petterson approximation is
not valid (a `low-viscosity' disc) for quiescent X-ray binaries.

Models of such low-viscosity discs using the Pringle--Ogilvie
equations of motion exhibit remarkable behavior: for a given obliquity
(angle between the black-hole spin axis and companion orbital axis)
there is {\em no} steady-state solution for $\beta$ smaller than some
critical value. We have argued at the end of \S\ref{sec:critical} that
the failure of these equations probably arises because they do not
allow hyperbolic behavior but the question of how warped 
low-viscosity Lense--Thirring discs actually behave remains to be
answered. 

The behavior of warped accretion discs around massive black holes is
equally rich. Here there is no significant companion torque (unless
the black hole is a member of a binary system), but the
Bardeen--Petterson approximation remains suspect because it neglects
the self-gravity of the disc. In fact we find that most plausible
models of AGN accretion discs have low viscosity in the sense that
viscous torques are smaller at all radii than one or both of the Lense--Thirring and
self-gravity torques.  If the viscosity is sufficiently small, spiral
bending waves are excited at the warp radius and propagate inward with
growing amplitude until they are eventually damped by viscosity or
non-linear effects. The presence of such waves may contribute to
obscuration of the disc and the illumination of the warped disc by the
central source may affect the disc spectrum or apparent size at
optical wavelengths. 

It is worth re-emphasizing that many of our conclusions are
based on a simple model of the internal stresses in the disc -- the
stress tensor is that of a viscous fluid and the viscosity is related
to the pressure through the Shakura--Sunyaev $\alpha$ parameter -- that
does not correspond to the actual stress tensor, which probably arises
mostly from anisotropic MHD turbulence. The available evidence on the
validity of this model from numerical MHD simulations, discussed at
the end of \S\ref{sec:visc}, suggests that it overestimates the rate
of viscous damping of warps; if correct, this would strengthen our
conclusions about the limited validity of the Bardeen--Petterson
approximation and the importance of tidal torques and self-gravity in
shaping warped accretion discs.

Our results suggest several avenues for future work. A better
treatment of self-gravitating warped discs would merge the
Pringle--Ogilvie equations (\ref{eq:ogthree}) with a description of
the mutual torques due to self-gravity as in
\cite{ger09}. Generalizing the Pringle--Ogilvie equations to include
wavelike behavior is also a necessary step for a complete description
of warped accretion discs. Understanding the actual behavior of
low-viscosity Lense--Thirring discs that exceed the critical obliquity
is important and challenging. Simple models of the emission from
warped discs may help to resolve current discrepancies between simple
flat $\alpha$-disc models and observations of AGN spectra and sizes.

We thank Julian Krolik, Jerome Orosz, and Jihad Touma for illuminating
discussions. We thank Gordon Ogilvie for many insights and for
providing the program used to calculate the viscosity coefficients
$Q_i$.  ST thanks the Max Planck Institute for Astrophysics and the
Alexander von Humboldt Foundation for hospitality and support during a
portion of this work. This research was supported in part by NASA grant
NNX11AF29G.

\end{document}